\documentclass[conference]{IEEEtran}
\usepackage{epsfig}
\usepackage{times}
\usepackage{float}
\usepackage{afterpage}
\usepackage{amsmath}
\usepackage{amstext}
\usepackage{amssymb,bm}
\usepackage{latexsym}
\usepackage{color}
\usepackage{graphicx}
\usepackage{amsmath}
\usepackage{amsthm}
\usepackage{graphicx}
\usepackage[center]{caption}
\usepackage{pstricks}
\usepackage{subfigure}
\usepackage{booktabs}
\usepackage{multicol}
\usepackage{lipsum}
\usepackage{dblfloatfix}
\allowdisplaybreaks

\newtheorem{thm}{Theorem}

\newtheorem{lem}[thm]{Lemma}

\newtheorem{rem}{Remark}

\begin{document}

\title{On Network Simplification for Gaussian Half-Duplex Diamond Networks}

\author{
\IEEEauthorblockN{Martina Cardone$^{\dagger}$, Christina Fragouli$^{\dagger}$, Daniela Tuninetti$^*$}
$^{\dagger}$ UCLA, Los Angeles, CA 90095, USA,
Email: \{martina.cardone, christina.fragouli\}@ucla.edu\\
$^*$ University of Illinois at Chicago,
Chicago, IL 60607, USA, 
Email: danielat@uic.edu
\thanks{
The work of M. Cardone and C. Fragouli was partially funded by NSF under award number 1514531.
The work of D. Tuninetti was partially funded by NSF under award number 1218635. 
M. Cardone would like to acknowledge insightful discussions with Yahya H. Ezzeldin.}
}

\IEEEoverridecommandlockouts
\maketitle
\begin{abstract}
This paper investigates the simplification problem in Gaussian Half-Duplex (HD) diamond networks. The goal is to answer the following question: what is the minimum (worst-case) fraction of the total HD capacity that one can always achieve by smartly selecting a subset of $k$ relays, out of the $N$ possible ones? We make progress on this problem for $k=1$ and $k=2$ and show that for $N=k+1, \ k \in \{1,2\}$ at least $\frac{k}{k+1}$ of the total HD capacity is always {approximately (i.e., up to a constant gap)} achieved. Interestingly, and differently from the Full-Duplex (FD) case, the ratio in HD depends on $N$, and decreases as $N$ increases. For all values of $N$ and $k$ for which we derive worst case fractions, we also show these to be {approximately} tight. This is accomplished by presenting $N$-relay Gaussian HD diamond networks for which the best $k$-relay subnetwork has {an approximate} HD capacity equal to the worst-case fraction of the total {approximate} HD capacity. Moreover, we provide additional comparisons between the performance of this simplification problem for HD and FD networks, which highlight their different natures.
\end{abstract}

\section{Introduction}
Wireless network simplification, first introduced in~\cite{NazarogluIT2014} in the context of Gaussian Full-Duplex (FD) diamond networks\footnote{An $N$-relay diamond network is a relay network topology where the source can communicate with the destination only through $N$ non-interfering relays.}, shows the surprising result that a significant fraction of the capacity can be achieved by using only a small subset of the available relays.
In this paper, we seek to answer the following question: how do these results extend to Gaussian Half-Duplex (HD) diamond networks?

The wireless simplification approach offers a number of benefits.
First, it promises {\it energy savings} since only the power of the active relays is used to transmit information; the power available at the relays that are kept silent is saved. 
Then, it simplifies the {\it synchronization} problem as only the selected relays have to be synchronized for transmission. 
Finally, for HD networks, simplification offers benefits in terms of {\it scheduling}.
Indeed, in an $N$-relay HD network, a capacity-achieving scheme requires a {\it global} optimization over the $2^N$ possible listen/transmit configuration states.
This approach, as $N$ increases, quickly becomes computationally prohibitive.
Thus, selecting a small subset of the relays leads to a significant complexity reduction in the scheduling.

In this paper, motivated by the numerous benefits of wireless simplification, we seek to understand how much of the HD capacity one can achieve by smartly selecting a subset of $k$ relays out of the $N$ possible ones in a Gaussian HD diamond network.
As a first step in this direction, we here provide a worst-case (i.e., independent of the channel parameters) {approximate (i.e., up to a constant gap)} capacity guarantee in terms of achievable fraction for $k=1$ and $k=2$ when $N=k+1$ and $N \gg 1$.
We also present network examples for which the best $k$-relay subnetwork {approximately} achieves the worst-case fraction {of the total HD capacity}, hence showing that the derived worst-performance guarantees are indeed tight.

Moreover, we find significant differences of the wireless simplification problem for HD and FD networks. 
For example, (i) in HD the fraction of the achieved capacity depends on $N$ and decreases as $N$ increases,  
(ii) the worst-case networks in HD and FD are not necessarily the same and 
(iii) the best $k$-relay subnetworks in HD and FD might be different.
\smallskip
\noindent{\bf{Related Work.}} 
The capacity characterization of the Gaussian HD relay network is a long-standing open problem. 
The tightest upper bound on the capacity is the well-known cut-set upper bound. 
In~\cite{CardoneIT2014}, this bound was evaluated by using the approach first proposed in~\cite{KramerAllerton2004} and shown to be achievable to within $1.96 (N+2)$ bits per channel use (independently of the channel parameters), by noisy network coding~\cite{LimIT2011}. 

In general, the evaluation of the cut-set upper bound requires an optimization over
$2^N$ listen/transmit states.
Recently, in~\cite{CardoneITW2015} the authors proved a surprising result: at most $N+1$ states (out of the $2^N$ possible ones) suffice for capacity characterization (up to a constant gap) for a class of HD relay networks, which includes the Gaussian {noise network.}
However, the problem of finding which are the $N+1$ active states might still require the use of all the $N$ relays. 

This work presents partial results on the wireless simplification problem for the HD case, whose FD counterpart was solved in~\cite{NazarogluIT2014}.
In particular, in~\cite{NazarogluIT2014} it was shown that, by selecting $k$ relays (out of the $N$ possible ones), one can always approximately achieve at least a fraction $\frac{k}{k+1}$ of the capacity. 
This result was proved to be tight, i.e., there exist $N$-relay Gaussian FD diamond networks for which the best $k$-relay subnetwork {approximately} achieves this fraction {of the total FD capacity.} 
A polynomial-time algorithm to discover these high-capacity $k$-relay subnetworks was also proposed.
In~\cite{BrahmaISIT2014Relay} the authors proved that, by selecting $k=2$ relays out of the $N$ possible ones {in a Gaussian HD diamond network}, and by operating them only in a complementary fashion {(i.e., when one relay listens, the other transmits), at least $\frac{1}{2}$ of the total HD capacity is approximately achieved.}

From the result in~\cite{NazarogluIT2014}, it directly follows that in HD, by selecting $k$ relays, one can always {approximately} achieve a fraction $\frac{k}{2(k+1)}$ of the HD capacity of the whole network. 
This is accomplished by operating the subnetwork in only $2$ states (out of the $2^k$ possible ones) of equal duration: the first where all the $k$ relays listen {and} the second where all the $k$ relays transmit.
However, as we show in this paper this capacity guarantee is not tight in general.
Moreover, differently from the work in~\cite{BrahmaISIT2014Relay}, in this paper we do not restrict the selected relays to operate only in certain states. 
This leads to better performance guarantees in terms of achievable fraction.

\smallskip
\noindent{\bf{Paper Organization.}}
Section~\ref{sec:SysModMainResu} describes the $N$-relay Gaussian HD diamond network and summarizes known capacity results.
Section~\ref{sec:MainProof} proves our main result, i.e., it provides a tight worst-case guarantee in terms of achievable fraction of the capacity for $k=1$ and $k=2$ when $N=k+1$ and $N \gg 1$. Section~\ref{sec:MainProof} also highlights differences between the simplification problem in HD and FD networks.
Finally, Section~\ref{sec:Concl} concludes the paper and discusses future research directions. 
Some of the proofs can be found in the Appendix.

\section{System Model and Known Results}
\label{sec:SysModMainResu}
With $\left [ n_1:n_2\right ]$ we denote the set of integers from $n_1$ to $n_2 \!\geq \!n_1$;
$\lfloor a \rfloor$ and $\lceil a \rceil$ are the floor and ceiling functions of $a \in \mathbb{R}$, respectively. 
Calligraphic letters denote sets; $|\mathcal{A}|$ is the cardinality of $\mathcal{A}$, $\mathcal{A} \backslash \mathcal{B}$ is the set of elements that belong to $\mathcal{A}$ but not to $\mathcal{B}$ and $\mathcal{A}^c$ is the complement of $\mathcal{A}$.
$\mathcal{A} \!\subseteq\! \mathcal{B}$ indicates that $\mathcal{A}$ is a subset of $\mathcal{B}$, $\mathcal{A} \!\cup \!\mathcal{B}$ is the union of $\mathcal{A}$ and $\mathcal{B}$ and $\mathcal{A}\! \cap\! \mathcal{B}$ is the intersection of $\mathcal{A}$ and $\mathcal{B}$; $\emptyset$ is the empty set.

The Gaussian HD diamond network consists of a source communicating with a destination only through $N$ non-interfering relay stations operating in HD. 
The input/output relationship for this network is
\begin{subequations}
\label{eq:inputoutput}
\begin{align}
Y_i & = \left (1-S_i \right ) h_{i} X_0 + Z_i, \ \forall i \in [1:N],
\\
Y_{N+1} & = \sum_{i=1}^N S_i g_{i} X_i + Z_{N+1},
\end{align}
\end{subequations}
where: 
(i) $X_0$ denotes the channel input at the source and $Y_{N+1}$ the channel output at the destination;
(ii) the channel inputs are subject to the average power constraint $\mathbb{E} \left [ |X_i|^2\right ]\leq 1,i \in [0:N]$;
(iii) $S_i, i \in [1:N]$, is the binary random variable, which represents the state (either listening or transmitting) of the $i$-th relay, with $i \in [1:N]$, i.e., if $S_i=0$ then it is receiving, while if $S_i=1$ then it is transmitting~\cite{KramerAllerton2004};
(iv) the channel parameters $\left(h_{i},g_{i} \right) \in \mathbb{C}^2, \forall i \in [1:N]$, are constant and therefore known to all terminals;
(v) the noises are independent.
In the following we shall indicate
\begin{subequations}
\begin{align}
\ell_i &:= \log(1+| h_{i}|^2), \ \forall i \in [1:N],
\\
r_i &:= \log(1+| g_{i}|^2), \ \forall i \in [1:N].
\end{align}
\end{subequations}
The {capacity}\footnote{We use standard definitions for codes, achievable rates and capacity.} $\bar{\mathsf{C}}^{\rm{HD}}$ of the channel in~\eqref{eq:inputoutput} is not known, but can be upper and lower bounded as
\begin{subequations}
\label{eq:capcutsetup}
\begin{align}
&{\mathsf{C}}^{{\rm{HD}}}_{\mathcal{N}_{\rm{F}}} - \mathsf{G}_1 \leq \bar{\mathsf{C}}^{\rm{HD}} \leq {\mathsf{C}}^{{\rm{HD}}}_{\mathcal{N}_{\rm{F}}} + \mathsf{G}_2,
\\
{\mathsf{C}}^{{\rm{HD}}}_{\mathcal{N}_{\rm{F}}} &:=\! \max_{\lambda} \!\!\min_{\mathcal{A}_{\rm{F}} \subseteq \mathcal{N}_{\rm{F}}} \!\!\!\sum_{s\in [0:1]^N} \!\!\lambda_s \!\left (\max_{i \in \mathcal{A}_{\rm{F}} \cap \mathcal{R}_s} \!\!\ell_i \!+\! \!\max_{i \in \mathcal{A}^c_{\rm{F}} \cap \mathcal{R}^c_s} \!\!r_i \right ),
 \label{eq:capApprox}
\end{align}
\end{subequations}
where:
(i) $\mathsf{G}_1$ and $\mathsf{G}_2$ are both ${O}(N)$ and independent of the actual value of the channel parameters;
(ii) $\lambda= \left[\lambda_s\right ]$ with $\lambda_{s} := \mathbb{P}[S_{[1: N]}=s] \in[0,1] : \sum_{s\in [0:1]^N} \lambda_{s}=1$;
(iii) $\mathcal{R}_s$ 
contains the relays that, in state $s\in [0:1]^N$, are receiving, i.e., among the relays `on the side of the destination' (indexed by $\mathcal{A}_{\rm{F}}$) only those in receive mode matter, and similarly, among the relays `on the side of the source' (indexed by $\mathcal{A}_{\rm{F}}^c$) only those in transmit mode matter.

In the next section we will use {the result in~\eqref{eq:capcutsetup}} to prove that a significant fraction of ${\mathsf{C}}^{{\rm{HD}}}_{\mathcal{N}_{\rm{F}}}$ can {always} be achieved by selecting $k\!=\!1$ and $k\!=\!2$ relays out of the $N$ possible ones.

\section{Achieving a fraction of the HD capacity}
\label{sec:MainProof}
In this section we prove our main result, i.e., we derive a worst-case fraction guarantee for any $N$-relay Gaussian HD diamond network with $k=1$ and $k=2$ when $N=k+1$ and {when} $N \gg 1$. 
Moreover, for each case we provide network examples which {approximately} achieve the derived fraction {of the total HD capacity}, hence showing that our worst-case performance guarantees are indeed tight.
Our main result is presented in the following theorem.

\begin{thm}
\label{th:theoremDiamHD}
In the Gaussian HD $N$-relay diamond network, by selecting $k \leq N$ relays and by keeping the remaining $N-k$ ones silent, we can achieve, up to a gap, a rate $\mathsf{C}^{{\rm{HD}}}_{k,N}$ such that
\begin{align}
\label{eq:PerfGarHD}
\frac{\mathsf{C}^{{\rm{HD}}}_{k,N}}{{\mathsf{C}}^{{\rm{HD}}}_{\mathcal{N}_{\rm{F}}}} \geq
\left \{
\begin{array}{ll}
\frac{1}{2} & k=1, \ N=2
\\
\frac{1}{4} & k=1, \ N \gg 1
\\
\frac{2}{3} & k=2, \ N=3
\\
\frac{1}{2}  & k=2, \ N \gg 1
\end{array}
\right..
\end{align}
Moreover, the bound in~\eqref{eq:PerfGarHD} is tight
up to a constant gap.
\end{thm}

Before going into the technical details of the proof of Theorem~\ref{th:theoremDiamHD} we make a couple of remarks.

\begin{rem}
\label{rem:DecrFrac}
Theorem~\ref{th:theoremDiamHD} shows that, for a fixed value of $k \in [1:2]$, the fraction $\frac{\mathsf{C}^{{\rm{HD}}}_{k,N} }{{\mathsf{C}}^{{\rm{HD}}}_{\mathcal{N}_{\rm{F}}}}$ decreases as $N$ increases.
In particular, for $N=k+1$ we have $\frac{\mathsf{C}^{{\rm{HD}}}_{k,N} }{{\mathsf{C}}^{{\rm{HD}}}_{\mathcal{N}_{\rm{F}}}} = \frac{k}{k+1}$ as in FD~\cite[Theorem 1]{NazarogluIT2014}.
Differently from FD where the fraction does not depend on $N$, in HD the fraction decreases as $N$ increases.
\end{rem}

\begin{rem}
The lower bound on $\mathsf{C}^{{\rm{HD}}}_{k,N}$ for $k=2$ and $N \gg 1$ is the one derived in~\cite[Theorem 2.2]{BrahmaISIT2014Relay} where the $k=2$ selected relays are constrained to operate in a complementary fashion. 
However, for $k=2$ and $N=3$, the result in Theorem~\ref{th:theoremDiamHD} improves over~\cite[Theorem 2.2]{BrahmaISIT2014Relay}, hence showing the importance of optimizing over all possible states.
\end{rem}

In what follows we let ${\mathsf{C}}^{\star{\rm{HD}}}_{\mathcal{N}_{i}}, i \in [1:2]$ be the HD achievable rate of the subnetwork $\mathcal{N}_i$ when operated with the `natural' schedule derived from $\lambda^{\star}$ (the optimal schedule of the whole network).
Clearly we have ${\mathsf{C}}^{\star{\rm{HD}}}_{\mathcal{N}_{i}} \leq {\mathsf{C}}^{{\rm{HD}}}_{\mathcal{N}_{i}}$.
The proof of Theorem~\ref{th:theoremDiamHD} makes use of the following lemma, whose proof can be found in Appendix~\ref{app:lemmaPart}.

\begin{lem}
\label{lem:part}
For any Gaussian HD diamond network $\mathcal{N}_{\rm{F}}$ with $N$ relays, we have
\begin{align}
{\mathsf{C}}^{{\rm{HD}}}_{\mathcal{N}_{\rm{F}}}
 \leq{\mathsf{C}}^{\star{\rm{HD}}}_{\mathcal{N}_{1}} +{\mathsf{C}}^{\star{\rm{HD}}}_{\mathcal{N}_{2}},
\end{align}
where $\left \{\mathcal{N}_1,\mathcal{N}_2 \right \}$ is a partition of the full network $\mathcal{N}_{\rm{F}} \!=\! [1:N]$.
\end{lem}


\subsection{The case $k=1$ and $N=2$}
From the result in Lemma~\ref{lem:part} we obtain
\begin{align*}
{\mathsf{C}}^{{\rm{HD}}}_{\mathcal{N}_{\rm{F}}} \!=\! {\mathsf{C}}^{{\rm{HD}}}_{\{1,2\}} \!\leq\!
{\mathsf{C}}^{\star{\rm{HD}}}_{\{1\}} \!+\!
{\mathsf{C}}^{\star{\rm{HD}}}_{\{2\}}
\!\leq\!
{\mathsf{C}}^{{\rm{HD}}}_{\{1\}}\!+\!
{\mathsf{C}}^{{\rm{HD}}}_{\{2\}}
\!\leq\!
2 \! \max_{i \in [1:2]} \left \{ {\mathsf{C}}^{{\rm{HD}}}_{\{i\}}\right \} \!,
\end{align*}
which implies $\mathsf{C}^{{\rm{HD}}}_{1,2} \!\geq \!\frac{1}{2}{\mathsf{C}}^{{\rm{HD}}}_{\mathcal{N}_{\rm{F}}}$.
We now present a network with $N\!=\!2$ where, by selecting the best relay, we {approximately} achieve $\frac{1}{2}$ of the HD capacity of the whole network.
\\
{\bf{Example.}}
Let $\ell_i \!=\! r_i, i \in [1:2]$ and $\ell_1\!=\!\ell_2$. For this network from~\cite{BagheriIT2014} we have ${\mathsf{C}}^{{\rm{HD}}}_{\mathcal{N}_{\rm{F}}} \!=\! \ell_1$ and $\mathsf{C}^{{\rm{HD}}}_{1,2}\!=\! \frac{\ell_1}{2}$, i.e., $\mathsf{C}^{{\rm{HD}}}_{1,2} \!=\! \frac{1}{2} {\mathsf{C}}^{{\rm{HD}}}_{\mathcal{N}_{\rm{F}}}$.

\subsection{The case $k=1$ and $N \gg 1$} 
From the result in~\cite[Theorem 1]{NazarogluIT2014} for $k=1$ we obtain
\begin{align*}
\frac{1}{2}{\mathsf{C}}^{{\rm{HD}}}_{\mathcal{N}_{\rm{F}}} \leq 
\frac{1}{2}{\mathsf{C}}^{{\rm{FD}}}_{\mathcal{N}_{\rm{F}}} \leq
\mathsf{C}^{{\rm{FD}}}_{1},
\end{align*} 
where we used the notation (i) $\mathsf{C}^{{\rm{FD}}}_{1}$ (which indicates the {approximate} FD capacity of the $k=1$ selected relay) to highlight that in FD the ratio $\frac{\mathsf{C}^{{\rm{FD}}}_{1}}{{\mathsf{C}}^{{\rm{FD}}}_{\mathcal{N}_{\rm{F}}}}$ does not depend on $N$ and (ii) ${\mathsf{C}}^{{\rm{FD}}}_{\mathcal{N}_{\rm{F}}}$ to indicate the approximate {FD capacity of the whole $2$-relay network}.
It is not difficult to see that $\mathsf{C}^{{\rm{FD}}}_{1} = 2 \tilde{\mathsf{C}}^{{\rm{HD}}}_{1,N}$, where $\tilde{\mathsf{C}}^{{\rm{HD}}}_{1,N}$ is the {approximate} HD capacity of the $k=1$ selected relay when it receives for $\frac{1}{2}$ of the time and it transmits for $\frac{1}{2}$ of the time. Thus,
\begin{align*}
\frac{1}{2}{\mathsf{C}}^{{\rm{HD}}}_{\mathcal{N}_{\rm{F}}} \!\leq\! 
\frac{1}{2}{\mathsf{C}}^{{\rm{FD}}}_{\mathcal{N}_{\rm{F}}} \!\leq\!
\mathsf{C}^{{\rm{FD}}}_{1}\!=\!
2 \tilde{\mathsf{C}}^{{\rm{HD}}}_{1,N} \leq
2\mathsf{C}^{{\rm{HD}}}_{1,N}
\Longrightarrow
\mathsf{C}^{{\rm{HD}}}_{1,N}
\!\geq \!\frac{1}{4} {\mathsf{C}}^{{\rm{HD}}}_{\mathcal{N}_{\rm{F}}}.
\end{align*}
We next provide a network example for which {(up to a constant gap)} $\mathsf{C}^{{\rm{HD}}}_{1,N}= \frac{1}{4}{\mathsf{C}}^{{\rm{HD}}}_{\mathcal{N}_{\rm{F}}}$ for $N \gg 1$.
\\
{\bf{Example.}}
We let
\begin{subequations}
\label{eq:worstrelayHDex}
\begin{align}
&\ell_i = \frac{\left \lceil \frac{N+2}{2} \right \rceil \left \lfloor \frac{N+2}{2} \right \rfloor}{\left \lceil \frac{N+2}{2} \right \rceil +\left \lfloor \frac{N+2}{2} \right \rfloor}\frac{\mathsf{c}}{i}, \ i \in [1:N],
\\ &r_i = \frac{\left \lceil \frac{N+2}{2} \right \rceil \left \lfloor \frac{N+2}{2} \right \rfloor}{\left \lceil \frac{N+2}{2} \right \rceil +\left \lfloor \frac{N+2}{2} \right \rfloor} \frac{\mathsf{c}}{N-i+1}, \ i \in [1:N],
\end{align}
\end{subequations}
where $\mathsf{c} \in \mathbb{R}_+$.
Notice that in this network we have $\ell_1 \geq \ell_2 \geq \ldots \geq \ell_N$ and $r_1 \leq r_2 \leq \ldots \leq r_N$. 
Moreover, it is not difficult to see that the vector of the relay-destination capacities is just the flipped version of the vector of the source-relay capacities. 
For the network in~\eqref{eq:worstrelayHDex} all the single-relay capacities are the same and, up to a constant gap, evaluate to
\begin{align}
\label{eq:singlecapHD}
\mathsf{C}^{{\rm{HD}}}_{1,N} = \frac{\left \lceil \frac{N+2}{2} \right \rceil \left \lfloor \frac{N+2}{2} \right \rfloor}{\left \lceil \frac{N+2}{2} \right \rceil +\left \lfloor \frac{N+2}{2} \right \rfloor} \frac{\mathsf{c}}{N+1}
\stackrel{N \gg 1}{=} \frac{\mathsf{c}}{4}.
\end{align}
\begin{figure}
\begin{center}
\includegraphics[width=0.9\columnwidth]{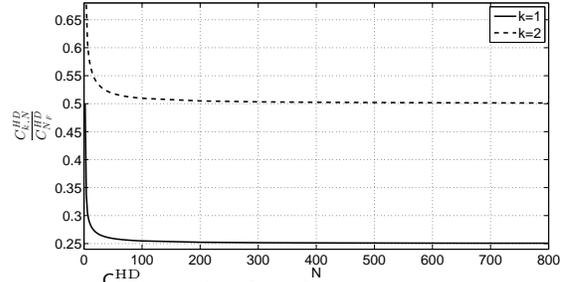}
\end{center}
\vspace{-6mm}
\caption{$\frac{\mathsf{C}^{{\rm{HD}}}_{k,N}}{{\mathsf{C}}^{{\rm{HD}}}_{\mathcal{N}_{\rm{F}}}}$ for $k\in [1:2]$ versus different values of $N \in [1:800]$ for the network in \eqref{eq:worstrelayHDex} with $\mathsf{c}=1$.}
\label{fig:fractionkrelayHDupper1}
\vspace{-2mm}
\end{figure}
For the network in~\eqref{eq:worstrelayHDex} in Fig.~\ref{fig:fractionkrelayHDupper1} we plot $\frac{\mathsf{C}^{{\rm{HD}}}_{1,N}}{{\mathsf{C}}^{{\rm{HD}}}_{\mathcal{N}_{\rm{F}}}}$ (see solid line) versus different values of $N \in [1:800]$ for $\mathsf{c}=1$. 
From Fig.~\ref{fig:fractionkrelayHDupper1} we observe that for $N \gg 1$ we have $\mathsf{C}^{{\rm{HD}}}_{1,N} = \frac{1}{4} {\mathsf{C}}^{{\rm{HD}}}_{\mathcal{N}_{\rm{F}}}$, i.e., the bound in Theorem~\ref{th:theoremDiamHD} for $k=1$ and $N \gg 1$ is precisely met. 
In particular, for $N \gg 1$, we have ${\mathsf{C}}^{{\rm{HD}}}_{\mathcal{N}_{\rm{F}}} = {\mathsf{C}}^{{\rm{FD}}}_{\mathcal{N}_{\rm{F}}}= \mathsf{c}$.

\subsection{The case $k=2$ and $N=3$}
{We here prove} that in every Gaussian HD diamond network with $N=3$ relays, there always exists a subnetwork of $k=2$ relays such that its {approximate} HD capacity is lower bounded by $\frac{2}{3}$ of the {approximate} HD capacity of the whole network. 
{To this end, we make use of the results in Lemma~\ref{lem:part}, Lemma~\ref{lem:existence} (whose proof can be found in Appendix~\ref{app:lemExist}) and Lemma~\ref{lem:MinCut}.

\begin{lem}
\label{lem:existence}
Consider a Gaussian HD diamond network $\mathcal{N}_{\rm{F}}$ with $N=3$. 
Let $\mathcal{A}^{\star}_i \subseteq \mathcal{N}_i = \left \{i,k \right \}$ with $\left | \mathcal{A}^{\star}_i \right |=a^\star_i$ and $\mathcal{A}^{\star}_j \subseteq \mathcal{N}_j = \left \{ i,j\right \}$ with $\left | \mathcal{A}^{\star}_j \right |=a^\star_j$ be the min cuts for the HD networks $\mathcal{N}_i$ and $\mathcal{N}_j$, respectively, so that $\left | a^\star_i-a^\star_j\right | \neq 2$.
Let the networks $\mathcal{N}_i$ and $\mathcal{N}_j$ operate with the optimal schedule $\lambda^{\star}$ of the whole network ${\mathsf{C}}^{{\rm{HD}}}_{\mathcal{N}_{\rm{F}}} $ with $N=3$ relays.
{Then, with ${\mathsf{C}}^{{\rm{HD}}}_{\mathcal{N}_{\rm{F}}} = {\mathsf{C}}^{{\rm{HD}}}_{\{1,2,3\}}$ we have}
\begin{align}
\label{eq:niceCond}
{\mathsf{C}}^{{\rm{HD}}}_{\mathcal{N}_{\rm{F}}} + {\mathsf{C}}^{\star{\rm{HD}}}_{\{i\}} \leq {\mathsf{C}}^{\star{\rm{HD}}}_{\{i,j\}} + {\mathsf{C}}^{\star{\rm{HD}}}_{\{i,k\}}.
\end{align}
\end{lem}

\begin{lem}
\label{lem:MinCut}
In any $3$-relay Gaussian diamond network (both FD and HD) let $\mathcal{A}^{\star}_1 \!\subseteq\! \mathcal{N}_1 \!=\! \{1,2\}$, $\mathcal{A}^{\star}_2 \!\subseteq\! \mathcal{N}_2 \!=\! \{1,3\}$
and
$\mathcal{A}^{\star}_3 \!\subseteq \!\mathcal{N}_3 \!=\! \{2,3\}$
be the min cuts of the networks $\mathcal{N}_1$, $\mathcal{N}_2$ and $\mathcal{N}_3$, respectively with $\left |\mathcal{A}^{\star}_i \right |\!=\!a_i^{\star}, \forall i \in [1:3]$.
Then, $\exists \left( i,j\right ) \in [1:3]^2$ with $i \neq j$ for which $\mathcal{A}^{\star}_i$ and $\mathcal{A}^{\star}_j$ satisfy $\left | a_i^{\star}-a_j^{\star}\right | \neq 2$.
\end{lem}

\begin{IEEEproof}
Assume, without loss of generality, that $\left | a_1^{\star}-a_2^{\star}\right | = 2$, i.e., either $\left \{ \mathcal{A}^{\star}_1,\mathcal{A}^{\star}_2 \right \} = \left \{ \emptyset, \{1,3\}\right \}$ or
$\left \{ \mathcal{A}^{\star}_1,\mathcal{A}^{\star}_2 \right \} = \left \{ \{1,2\},\emptyset\right \}$.
Let $\left \{ \mathcal{A}^{\star}_1,\mathcal{A}^{\star}_2 \right \} = \left \{ \emptyset, \{1,3\}\right \}$ (the same reasoning also holds for $\left \{ \mathcal{A}^{\star}_1,\mathcal{A}^{\star}_2 \right \} = \left \{ \{1,2\},\emptyset\right \}$). Then,

\noindent $\bullet$ if $\mathcal{A}^{\star}_3 = \emptyset$, then $\left | a^{\star}_1 - a^{\star}_3\right | = 0 \neq 2$;

\noindent $\bullet$ if $\mathcal{A}^{\star}_3 \!=\! \left \{ 2 \right \}$, then $\left | a^{\star}_1 - a^{\star}_3\right | \!= \!1 \!\neq \!2$ and $\left | a^{\star}_2 \!-\! a^{\star}_3\right | \!=\! 1 \!\neq \!2$;

\noindent $\bullet$ if $\mathcal{A}^{\star}_3 \!=\! \left \{ 3 \right \}$, then $\left | a^{\star}_1 \!-\! a^{\star}_3\right | \!=\! 1 \neq 2$ and $\left | a^{\star}_2 \!-\! a^{\star}_3\right | \!=\! 1 \neq 2$;

%
\noindent $\bullet$ if $\mathcal{A}^{\star}_3 \!=\! \left \{ 2,3 \right \}$, then $\left | a^{\star}_2 \!-\! a^{\star}_3\right | \!=\! 0 \neq 2$.
\end{IEEEproof}
\smallskip
Thanks to the result in Lemma~\ref{lem:MinCut} we know that $\exists i \in [1:3]$ such that \eqref{eq:niceCond} is satisfied. Without loss of generality, let $i=1$.
Moreover, from Lemma~\ref{lem:part} we have
${\mathsf{C}}^{{\rm{HD}}}_{\mathcal{N}_{\rm{F}}} \leq {\mathsf{C}}^{\star{\rm{HD}}}_{\{1\}} + {\mathsf{C}}^{\star{\rm{HD}}}_{\{2,3\}}$.
By summing this with \eqref{eq:niceCond} evaluated in $i\!=\!1$ we obtain
$2{\mathsf{C}}^{{\rm{HD}}}_{\mathcal{N}_{\rm{F}}} \!\leq\! {\mathsf{C}}^{\star{\rm{HD}}}_{\{1,2\}}+{\mathsf{C}}^{\star{\rm{HD}}}_{\{1,3\}} + {\mathsf{C}}^{\star{\rm{HD}}}_{\{2,3\}} 
\!\leq\! {\mathsf{C}}^{{\rm{HD}}}_{\{1,2\}}+{\mathsf{C}}^{{\rm{HD}}}_{\{1,3\}} + {\mathsf{C}}^{{\rm{HD}}}_{\{2,3\}}
\!\leq 3\! \max \left \{{\mathsf{C}}^{{\rm{HD}}}_{\{1,2\}},{\mathsf{C}}^{{\rm{HD}}}_{\{1,3\}},{\mathsf{C}}^{{\rm{HD}}}_{\{2,3\}} \right \}$, which implies $\mathsf{C}^{{\rm{HD}}}_{2,3} \!\geq \!\frac{2}{3}{\mathsf{C}}^{{\rm{HD}}}_{\mathcal{N}_{\rm{F}}}$.
}
We now provide a network example with $N\!=\!3$ where, by selecting the best subnetwork of $k\!=\!2$ relays, we {approximately} achieve $\frac{2}{3}$ of the HD capacity of the whole network.
\\
{\bf{Example.}}
{Let
$\ell_1 \!=\! \frac{1}{3}, \ \ell_{[2:3]}\!=\! 1, \ \ r_1 \!=\! r, \ r_{[2:3]}\!=\! \frac{2}{3}$ with $r \! \rightarrow \!\infty$.}
For this network ${\mathsf{C}}^{{\rm{HD}}}_{\mathcal{N}_{\rm{F}}} \!=\!1$ and $\mathsf{C}^{{\rm{HD}}}_{2,3} \!=\! \frac{2}{3}$, thus $\mathsf{C}^{{\rm{HD}}}_{2,3} \!=\! \frac{2}{3} {\mathsf{C}}^{{\rm{HD}}}_{\mathcal{N}_{\rm{F}}}$.

\subsection{The case $k=2$ and $N \gg 1$}
The lower bound in Theorem~\ref{th:theoremDiamHD} for this case is simply the one derived in~\cite[Theorem 2.2]{BrahmaISIT2014Relay}, valid for the case when the $k=2$ selected relays are allowed to operate only in a complementary fashion.
We next provide a network example for which $\mathsf{C}^{{\rm{HD}}}_{2,N} \!= \!\frac{1}{2} {\mathsf{C}}^{{\rm{HD}}}_{\mathcal{N}_{\rm{F}}}$ for $N \!\gg\! 1$, up to a constant gap.

\noindent {\bf{Example.}} We again consider the network defined in \eqref{eq:worstrelayHDex}.
Since for this network all the single-relay capacities, given in~\eqref{eq:singlecapHD}, are the same, then $\mathsf{C}^{{\rm{HD}}}_{2,N} \leq 2 \mathsf{C}^{{\rm{HD}}}_{1,N}$. 
It is not difficult to see that by selecting the first and the last relays we exactly get 
\begin{align*}
\mathsf{C}^{{\rm{HD}}}_{2,N} = 2\mathsf{C}^{{\rm{HD}}}_{1,N} = 2 \frac{\left \lceil \frac{N+2}{2} \right \rceil \left \lfloor \frac{N+2}{2} \right \rfloor}{\left \lceil \frac{N+2}{2} \right \rceil +\left \lfloor \frac{N+2}{2} \right \rfloor} \frac{\mathsf{c}}{N+1} \stackrel{N \gg 1}{=} \frac{\mathsf{c}}{2}.
\end{align*}
This, with the fact that for $N \gg 1$ we have ${\mathsf{C}}^{{\rm{HD}}}_{\mathcal{N}_{\rm{F}}} = \mathsf{c}$, gives the claimed ratio as we observe from Fig.~\ref{fig:fractionkrelayHDupper1} (dashed line).

\subsection{HD versus FD simplification}
We here discuss differences between the selection performances in HD and FD networks. 
We believe that the reason for this different behavior is that in HD the schedule plays a key role: removing some of the relays might change the global schedule of the network.

\noindent
$\bullet$ {\bf{In HD the ratio $\frac{\mathsf{C}^{{\rm{HD}}}_{k,N} }{{\mathsf{C}}^{{\rm{HD}}}_{\mathcal{N}_{\rm{F}}}}$ decreases as $N$ increases.}}
As already highlighted in Remark~\ref{rem:DecrFrac} this represents a surprising difference with respect to FD and shows that FD and HD relay networks have a different nature.

\noindent
$\bullet$ {\bf{Worst-case networks in HD and FD are not necessarily the same.}}
Consider the network example $\ell_1 \!=\! \ell_2 \!=\! r_1 \!=\! r_2 \!=\! 1$ and suppose we want to select $k\!=\!1$ relay.
We already showed before that, by selecting either the first or the second relay, we get $\mathsf{C}^{{\rm{HD}}}_{1,2} \!=\! \frac{1}{2}{\mathsf{C}}^{{\rm{HD}}}_{\mathcal{N}_{\rm{F}}}$.
Now, suppose we operate this network in FD. 
Then, it is not difficult to see that, by selecting any of the $N\!=\!2$ relays, we get $\mathsf{C}^{{\rm{FD}}}_{1} = {\mathsf{C}}^{{\rm{FD}}}_{\mathcal{N}_{\rm{F}}}$, which is greater than $\frac{1}{2}$, i.e., the worst-case ratio proved in~\cite[Theorem 1]{NazarogluIT2014}.

\noindent
$\bullet$ {\bf{The best HD and FD subnetworks are not necessarily the same.}}
Consider the network in~\eqref{eq:worstrelayHDex}. 
When the $N$ relays operate in HD, they all have the same single capacity given in~\eqref{eq:singlecapHD}. 
This means that by selecting any of the relays (i.e., at random) we get the same performance guarantee. 
Differently, when the $N$ relays operate in FD, only the $\lfloor \frac{N+2}{2} \rfloor$-th relay  (when $N$ is an odd number) and the relays number $\lfloor \frac{N+2}{2} \rfloor$ and number $\lfloor \frac{N+2}{2} \rfloor-1$ (when $N$ is an even number) give the performance guarantee of~\cite[Theorem 1]{NazarogluIT2014}.
As another example consider a Gaussian $2$-relay diamond network with $\ell_1=1$, $\ell_2=\frac{2}{5}$, $r_1=\frac{1}{2}$ and $r_2=\frac{14}{5}$ and suppose we want to select the best relay. 
It is not difficult to see that if the relays operate in FD, then the first relay is the best and it achieves $\mathsf{C}^{{\rm{FD}}}_{1} =\frac{1}{2}$, while if the relays operate in HD then the second relay is the best giving $\mathsf{C}^{{\rm{HD}}}_{1,2} =\frac{7}{20}$.


\noindent
$\bullet$ {\bf{Worst-case ratio with respect to FD, i.e., $\frac{\mathsf{C}^{{\rm{HD}}}_{k,N}}{{\mathsf{C}}^{{\rm{FD}}}_{\mathcal{N}_{\rm{F}}}}$}.}
The problem of finding the capacity of HD relay networks is computationally expensive, as it requires an optimization over $2^N$ cuts each of which depends on $2^N$ listen/transmit states.
On the contrary, the cut-set upper bound in FD can be more easily evaluated.
Thus, one can think of comparing the HD capacity of the $k$-relay selected subnetwork with respect to the FD capacity of the whole network.
By doing so, we get the results presented in the next theorem, whose proof can be found in Appendix~\ref{app:AppFDHD}.
\begin{thm}
\label{th:theoremDiamFD}
In the Gaussian HD $N$-relay diamond network, by selecting $k \leq N$ relays and by keeping the remaining $N-k$ ones silent, we can achieve, up to a gap, a rate $\mathsf{C}^{{\rm{HD}}}_{k,N}$ such that
\begin{align}
\label{eq:PerfGarFD}
\frac{\mathsf{C}^{{\rm{HD}}}_{k,N}}{{\mathsf{C}}^{{\rm{FD}}}_{\mathcal{N}_{\rm{F}}}} \geq
\left \{
\begin{array}{ll}
\frac{1}{2}  & k=1, \ N=1
\\
\frac{1}{3}   & k=1, \ N=2
\\
\frac{1}{4} & k=1, \ N \gg 1
\\
\frac{1}{2} & k=2, \ N \in \left [2:3 \right ]
\end{array}
\right. .
\end{align}
Moreover, the bound in~\eqref{eq:PerfGarFD} is tight
up to a constant gap.
\end{thm}

Although, for the case $k=2$, Theorem~\ref{th:theoremDiamFD} provides a lower bound for $N \in [2:3]$, we believe that $\mathsf{C}^{{\rm{HD}}}_{2,N} \geq \frac{1}{2}{\mathsf{C}}^{{\rm{FD}}}_{\mathcal{N}_{\rm{F}}}$ for $k=2$ and $N \geq 2$.
If this conjecture is proved to be true, then it would imply $\mathsf{C}^{{\rm{HD}}}_{k,N} \geq \frac{1}{2}{\mathsf{C}}^{{\rm{FD}}}_{\mathcal{N}_{\rm{F}}}$ for $k>1$ and $N \geq k$, as by selecting more relays (i.e., by increasing $k$) the performance guarantee cannot decrease.
By comparing Theorem~\ref{th:theoremDiamHD} and Theorem~\ref{th:theoremDiamFD}, two conclusions can be drawn:

\begin{enumerate}
\item For $N\!=\!k\!+\!1$, $k \!\in\! [1:2]$, Theorem~\ref{th:theoremDiamHD} provides a better guarantee with respect to Theorem~\ref{th:theoremDiamFD}, because of the use of a tighter upper bound to the HD performance.

\item For $N \gg 1$ and $k = 1$ the two bounds in Theorem~\ref{th:theoremDiamHD} and Theorem~\ref{th:theoremDiamFD} coincide.
This may indicate that, as $N$ increases, the schedule optimization is less crucial.
\end{enumerate}

\section{Conclusions and Future Work}
\label{sec:Concl}
We studied the simplification problem in an $N$-relay Gaussian HD diamond network.
We showed that, when $N=k+1$, by selecting $k \in [1:2]$ relays one can achieve (to within a constant gap) at least $\frac{k}{k+1}$ of the total HD capacity.
Differently from the FD case, this fraction decreases as $N$ increases.

The extension of these results to $k>2$ and the design of a polynomial-time algorithm which efficiently discovers a high-capacity $k$-relay subnetwork are interesting open problems, which are object of current investigation.

\appendices
\begin{table*}
\caption{Proof of Lemma \ref{prope:FDcut}.}
\begin{center}
 \begin{tabular}{||c c c c c c||} 
 \hline
 $\mathcal{A}_1$ & $\mathcal{A}_2$ & 
$f \left( \mathcal{A}_1,\mathcal{A}_2\right)$ & $\mathcal{A}_{\rm{F}}$ & $\mathcal{A}_{\rm{S}}$ & $g \left( \mathcal{A}_{\rm{F}},\mathcal{A}_{\rm{S}}\right)$  \\ [0.5ex] 
 \hline\hline
 $\emptyset$ & $\emptyset$ & 
$\max \left \{ r_i,r_j \right \} + \max \left \{ r_i,r_k \right \}$ & $\emptyset$ & $\emptyset$ &
$\max \left \{ r_i,r_j,r_k\right \} + r_i$ \\ 
 \hline
 $\emptyset$ & $\{i\}$ & 
$\max \left \{ r_i,r_j \right \} + \ell_i + r_k$ & $\emptyset$ & $\{i\}$ & $\max \left \{ r_i,r_j,r_k\right \} + \ell_i$ \\
 \hline
 $\emptyset$ & $\{k\}$ &  
$\max \left \{ r_i,r_j \right \} + \ell_k + r_i$ & $\{k \}$ & $\emptyset$ & $\ell_k + \max \left\{ r_i,r_j\right \} + r_i$ \\
 \hline
 $\{i\}$ & $\emptyset$ & 
$\ell_i + r_j + \max \left \{ r_i,r_k \right \}$ & $\emptyset$ & $\{ i\}$ & $\max \left \{ r_i,r_j,r_k\right \} + \ell_i$ \\
 \hline
 $\{i\}$ & $\{i\}$ & 
$2 \ell_i + r_j + r_k$ & $\{i \}$ & $\{i \}$ & $2\ell_i + \max \left \{ r_j, r_k\right \}$ \\ 
 \hline
 $\{i\}$ & $\{k\}$ & 
$\ell_i + r_j + \ell_k +r_i$ & $\{i,k\}$ & $\emptyset$ & $\max \left\{\ell_i, \ell_k \right \} + r_j + r_i$ \\ 
 \hline
 $\{i\}$ & $\{i,k\}$ & 
$\ell_i + r_j + \max \left \{ \ell_i, \ell_k\right \}$ & $\{i,k\}$ & $\{i\}$ & $\max \left \{ \ell_i, \ell_k\right \} + r_j + \ell_i$  \\ 
\hline
 $\{j\}$ & $\emptyset$ & 
$\ell_j + r_i + \max \left \{ r_i,r_k \right \}$ & $\{j\}$ & $\emptyset$ & $\ell_j + \max \left \{ r_i,r_k \right \} + r_i$ \\
\hline
 $\{j\}$ & $\{i\}$ & 
$\ell_j + r_i + \ell_i + r_k$ & $\{j\}$ & $\{i\}$ & $\ell_j + \max \left \{ r_i,r_k\right \} + \ell_i$ \\
\hline
 $\{j\}$ & $\{k\}$ & 
$\ell_j + 2 r_i + \ell_k$ & $\{j,k\}$ & $\emptyset$ & $\max \left \{ \ell_j, \ell_k \right \} + 2 r_i$ \\
\hline
 $\{j\}$ & $\{i,k\}$ & 
$\ell_j + r_i + \max \left \{\ell_i, \ell_k \right \}$ & $\{i,j,k\}$ & $\emptyset$ & $\max \left \{ \ell_i,\ell_j,\ell_k \right \} + r_i$  \\
\hline
 $\{i,j\}$ & $\{i,k\}$ & 
$\max \left \{ \ell_i,\ell_j \right \} + \max \left \{ \ell_i,\ell_k \right \}$ & $\{i,j,k\}$ & $\{i\}$ & $\max \left \{ \ell_i,\ell_j, \ell_k\right \} + \ell_i$ \\
[1ex] 
 \hline
\end{tabular}
\end{center}
\label{table:proofLemma7}
\end{table*}

\section{Proof of Lemma \ref{lem:part}}
\label{app:lemmaPart}
We here prove the result in Lemma~\ref{lem:part}. To this end we make use of the following lemma.
\begin{lem}
For any Gaussian FD diamond network, we have
\begin{align}
\mathsf{C}^{\rm{FD}}_{\mathcal{N}_{\rm{F}}} \leq {\mathsf{C}}^{\rm{FD}}_{\mathcal{N}_1} +{\mathsf{C}}^{\rm{FD}}_{\mathcal{N}_2},
\end{align}
where $\left \{\mathcal{N}_1,\mathcal{N}_2 \right \}$ is a partition of the full network $\mathcal{N}_{\rm{F}}$,
with
\begin{align*}
&{\mathsf{C}}^{\rm{FD}}_{\mathcal{N}_{\rm{F}}} 
= \min_{\mathcal{A}_{\rm{F}} \subseteq \mathcal{N}_{\rm{F}}} I_{\mathcal{A}_{\rm{F}}; \mathcal{N}_{\rm{F}}},
I_{\mathcal{A}_{\rm{F}}; \mathcal{N}_{\rm{F}}} 
:= \max_{i \in \mathcal{A}_{\rm{F}}} \ell_i + \max_{i \in \mathcal{N}_{\rm{F}} \backslash \mathcal{A}_{\rm{F}}} r_i,  
\\
& {\mathsf{C}}^{\rm{FD}}_{\mathcal{N}_j} 
\!=\! \min_{\mathcal{A}_j \!\subseteq\! \mathcal{N}_j} \!I_{\mathcal{A}_j; \mathcal{N}_j},
I_{\mathcal{A}_j; \mathcal{N}_j} 
\!:=\! \max_{i \in \mathcal{A}_j} \ell_i \!+\!\!\! \max_{i \in \mathcal{N}_j \backslash \mathcal{A}_j} \!\!r_i, j \!\in\! [1:2]. 
\end{align*}
\end{lem}

\begin{IEEEproof}
We have
\begin{align}
  &I_{\mathcal{A}_1; \mathcal{N}_1} \!+\! I_{\mathcal{A}_2;\mathcal{N}_2} \nonumber
= \max_{i \in \mathcal{A}_1} \ell_i\!+\!\max_{i \in \mathcal{A}_2} \ell_i \!+\! \max_{i \in \mathcal{N}_1 \backslash \mathcal{A}_1} r_i \!+\! \max_{i \in \mathcal{N}_2 \backslash \mathcal{A}_2} r_i  \nonumber
\\& \geq \max_{i \in \mathcal{A}_1 \cup \mathcal{A}_2} \ell_i+ \max_{i \in \left(\mathcal{N}_1 \backslash \mathcal{A}_1\right) \cup \left(\mathcal{N}_2 \backslash \mathcal{A}_2\right)} r_i \nonumber
\\&\stackrel{{\rm{(a)}}}{=} 
\max_{i \in \mathcal{A}_1 \cup \mathcal{A}_2} \ell_i+ \max_{i \in \left(\mathcal{N}_1 \cup \mathcal{N}_2 \right) \backslash{ \left(\mathcal{A}_1 \cup \mathcal{A}_2 \right)}} r_i \nonumber
\\&= I_{\mathcal{A}_1 \cup \mathcal{A}_2; {\mathcal{N}_{\rm{F}}}}
\geq \min_{\mathcal{B} \subseteq \mathcal{N}_{\rm{F}}} I_{\mathcal{B}; \mathcal{N}_{\rm{F}}} = \mathsf{C}_{\mathcal{N}_{\rm{F}}}^{\rm{FD}},
\label{eq:boundsFDpartition}
\end{align}
where the equality in $\rm{(a)}$ follows since $\mathcal{N}_1\! \cap\! \mathcal{A}_2 \!=\! \emptyset$ and $\mathcal{N}_2 \!\cap\! \mathcal{A}_1 \!=\! \emptyset$ and
$\left( \mathcal{B} \backslash \mathcal{A} \right) \!\cup\! \left( \mathcal{C} \backslash \mathcal{A} \right) \!=\! \left( \mathcal{B} \!\cup\! \mathcal{C}\right) \backslash \mathcal{A}$.
The result in \eqref{eq:boundsFDpartition} is valid $\forall \mathcal{A}_1 \!\subseteq\! \mathcal{N}_1$ and $\forall \mathcal{A}_2 \!\subseteq\! \mathcal{N}_2$, hence also for the minimum cuts of the networks $\mathcal{N}_1$ and $\mathcal{N}_2$, i.e., ${\mathsf{C}}^{\rm{FD}}_{\mathcal{N}_1} \!+\!{\mathsf{C}}^{\rm{FD}}_{\mathcal{N}_2} \!\geq\! \mathsf{C}_{\mathcal{N}_{\rm{F}}}^{\rm{FD}}$.
\end{IEEEproof}
\smallskip
We now show how the result in~\eqref{eq:boundsFDpartition} extends to HD.
We notice that ${\mathsf{C}}^{{\rm{HD}}}_{\mathcal{N}_{\rm{F}}}$
in~\eqref{eq:capApprox} can be equivalently written as
\begin{align*}
{\mathsf{C}}^{{\rm{HD}}}_{\mathcal{N}_{\rm{F}}}
 & = \min_{\mathcal{A}_{\rm{F}} \subseteq \mathcal{N}_{\rm{F}}} \sum_{s\in [0:1]^N} \!\!\lambda_s^{\star} \left( \max_{i \in\mathcal{A}_{\rm{F}} }  \ell_{i,s}^{\prime} + \max_{i \in \mathcal{A}_{\rm{F}}^c }  r_{i,s}^{\prime}
 \right),
\end{align*}
where
\begin{align}
\label{eq:NewChannels}
\ell_{i,s}^{\prime} \!=\! \left \{
\begin{array}{ll}
\ell_i & \text{if} \ i \in \mathcal{R}_s
\\
0 & \text{otherwise}
\end{array}
\right.,
\quad 
r_{i,s}^{\prime} \!=\! \left \{
\begin{array}{ll}
r_i & \text{if} \ i \in \mathcal{R}^c_s
\\
0 & \text{otherwise}
\end{array}
\right. .
\end{align}
From \eqref{eq:boundsFDpartition}, $\forall \mathcal{A}_1 \subseteq \mathcal{N}_1$ and $\forall \mathcal{A}_2 \subseteq \mathcal{N}_2$, with $\mathcal{N}_1,\mathcal{N}_2$ being a partition of $\mathcal{N}_{\rm{F}}$, we have that
\begin{align*}
\sum_{s\in [0:1]^N} \!\!\lambda_s^{\star}\left [ 
   \max_{i \in\mathcal{A}_1 }  \ell_{i,s}^{\prime}
 + \max_{i \in\mathcal{A}_2 }  \ell_{i,s}^{\prime}
 + \max_{i \in \mathcal{N}_1 \backslash \mathcal{A}_1 }  r_{i,s}^{\prime}
 + \max_{i \in \mathcal{N}_2 \backslash \mathcal{A}_2 }  r_{i,s}^{\prime}
\right ]
\\
\geq \sum_{s\in [0:1]^N} \!\!\lambda_s^{\star} \left( 
  \max_{i \in\mathcal{A}_{\rm{F}} }  \ell_{i,s}^{\prime}
+ \max_{i \in \mathcal{A}_{\rm{F}}^c }  r_{i,s}^{\prime} \right) \geq {\mathsf{C}}^{{\rm{HD}}}_{\mathcal{N}_{\rm{F}}},
\end{align*}
for $\mathcal{A}_1 \cup \mathcal{A}_2 = \mathcal{A}_{\rm{F}}$, which implies ${\mathsf{C}}^{{\rm{HD}}}_{\mathcal{N}_{\rm{F}}} \leq {\mathsf{C}}^{\star{\rm{HD}}}_{\mathcal{N}_{1}} +{\mathsf{C}}^{\star{\rm{HD}}}_{\mathcal{N}_{2}}$.

\section{Proof of Lemma \ref{lem:existence}}
\label{app:lemExist}
We here prove the result in Lemma~\ref{lem:existence}. To this end we make use of the following lemma.
\begin{lem}
\label{prope:FDcut}
Consider a FD diamond network $\mathcal{N}_{\rm{F}}$ with $N=3$.
Let $\mathcal{N}_1 = \left \{ i,j\right \}$
and $\mathcal{N}_2 = \left \{ i,k\right \}$ be two subnetworks of $\mathcal{N}_{\rm{F}}=\{1,2,3\}$ with $\left( i,j,k\right) \in [1:3]^3$ and $i \neq j \neq k$.
Let $\mathcal{A}_1 \subseteq \mathcal{N}_1$ with $\left | \mathcal{A}_1 \right | =a_1$, 
and $\mathcal{A}_2 \subseteq \mathcal{N}_2$ with $\left | \mathcal{A}_2 \right | =a_2$.
Define $\mathcal{N}_{\rm{S}} = \mathcal{N}_1 \cap \mathcal{N}_2$.
Clearly $\mathcal{N}_{\rm{F}} = \mathcal{N}_1 \cup \mathcal{N}_2$. 
If $\left |a_1-a_2\right | \neq 2$, then there exist $\mathcal{A}_{\rm{F}} \subseteq \mathcal{N}_{\rm{F}}$, $\mathcal{A}_{\rm{S}} \subseteq \mathcal{N}_{\rm{S}}$  such that
\begin{align}
\label{eq:propeFD}
&\max_{t \in \mathcal{A}_1} \ell_t + \max_{t \in \mathcal{N}_1 \backslash \mathcal{A}_1} r_t + \max_{t \in \mathcal{A}_2} \ell_t + \max_{t \in \mathcal{N}_2 \backslash \mathcal{A}_2} r_t \nonumber
\\&\geq
\max_{t \in \mathcal{A}_{\rm{F}}} \ell_t + \max_{t \in \mathcal{N}_{\rm{F}} \backslash \mathcal{A}_{\rm{F}}} r_t
+
\max_{t \in \mathcal{A}_{\rm{S}}} \ell_t + \max_{t \in \mathcal{N}_{\rm{S}} \backslash \mathcal{A}_{\rm{S}}} r_t.
\end{align}
\end{lem}

\begin{IEEEproof}
We define
\begin{align*}
f \left( \mathcal{A}_1,\mathcal{A}_2\right) &:= \max_{t \in \mathcal{A}_1} \ell_t + \max_{t \in \mathcal{N}_1 \backslash \mathcal{A}_1} r_t + \max_{t \in \mathcal{A}_2} \ell_t + \max_{t \in \mathcal{N}_2 \backslash \mathcal{A}_2} r_t,
\\ g \left( \mathcal{A}_{\rm{F}},\mathcal{A}_{\rm{S}}\right) &:=\max_{t \in \mathcal{A}_{\rm{F}}} \ell_t + \max_{t \in \mathcal{N}_{\rm{F}} \backslash \mathcal{A}_{\rm{F}}} r_t
+
\max_{t \in \mathcal{A}_{\rm{S}}} \ell_t + \max_{t \in \mathcal{N}_{\rm{S}} \backslash \mathcal{A}_{\rm{S}}} r_t.
\end{align*}
Table \ref{table:proofLemma7} proves that, $\forall \mathcal{A}_1 \subseteq \mathcal{N}_1$ and $\forall \mathcal{A}_2 \subseteq \mathcal{N}_2$ for which $\left |a_1-a_2\right | \neq 2$, then $f \left( \mathcal{A}_1,\mathcal{A}_2\right) \geq g \left( \mathcal{A}_{\rm{F}},\mathcal{A}_{\rm{S}}\right)$.

Notice that we did not consider the cases $\left \{\mathcal{A}_1, \mathcal{A}_2\right \} = \left \{\{i,j\}, \{i\} \right \}$ and $\left \{\mathcal{A}_1, \mathcal{A}_2\right \} = \left \{\{i,j\}, \{k\} \right \}$ since these are equivalent to $\left \{\mathcal{A}_1, \mathcal{A}_2\right \} = \left \{\{i\}, \{i,k\} \right \}$ and to $\left \{\mathcal{A}_1, \mathcal{A}_2\right \} = \left \{\{j\}, \{i,k\} \right \}$, respectively.
Therefore, Table \ref{table:proofLemma7} covers all the cases except $\left \{\mathcal{A}_1, \mathcal{A}_2 \right \} = \left \{ \emptyset, \{i,k\} \right \}$ and $\left \{\mathcal{A}_1, \mathcal{A}_2 \right \} = \left \{ \{i,j\},\emptyset \right \}$ since for these cases $\left |a_1-a_2\right | = 2$.
\end{IEEEproof}

We are now ready to prove the result in Lemma~\ref{lem:existence}.
Let, without loss of generality, $i=1$, $j=2$ and $k=3$, i.e., consider $\mathcal{A}^{\star}_1 \subseteq \mathcal{N}_1 = \left \{1,3 \right \}$ and $\mathcal{A}^{\star}_2 \subseteq \mathcal{N}_2 = \left \{ 1,2\right \}$ with $\left | a^\star_1-a^\star_2\right | \neq 2$.
With this we have
\begin{align*}
&{\mathsf{C}}^{\star{\rm{HD}}}_{\{1,2\}}+ {\mathsf{C}}^{\star{\rm{HD}}}_{\{1,3\}}
\\=& \!\!\sum_{s\in [0:1]^N} \!\!\!\!\lambda_s^{\star} \left ( \max_{i \in \mathcal{A}^{\star}_1} \ell_{i,s}^{\prime} \!+\! \max_{i \in \mathcal{N}_1 \backslash \mathcal{A}^{\star}_1} r_{i,s}^{\prime} \!+\!
\max_{i \in \mathcal{A}^{\star}_2} \ell_{i,s}^{\prime} \!+\! \max_{i \in \mathcal{N}_2 \backslash \mathcal{A}^{\star}_2} r_{i,s}^{\prime} 
\right )
\\  \stackrel{{\rm{(a)}}}{\geq} & \!\!\!\!
\sum_{s\in [0:1]^N}\! \!\!\!\lambda_s^{\star} \left ( \max_{i \in \mathcal{A}_{\rm{F}}} \ell_{i,s}^{\prime} \!+\! \max_{i \in \mathcal{N}_{\rm{F}} \backslash \mathcal{A}_{\rm{F}}} r_{i,s}^{\prime}
\!+\!
\max_{i \in \mathcal{A}_{\rm{S}}} \ell_{i,s}^{\prime} \!+\! \max_{i \in \mathcal{N}_{\rm{S}} \backslash \mathcal{A}_{\rm{S}}}r_{i,s}^{\prime}\right )
\\  = & \sum_{s\in [0:1]^N} \!\!\lambda_s^{\star} \left ( \max_{i \in \mathcal{A}_{\rm{F}}} \ell_{i,s}^{\prime} + \max_{i \in \mathcal{N}_{\rm{F}} \backslash \mathcal{A}_{\rm{F}}} r_{i,s}^{\prime}
\right ) 
\\&+ 
\sum_{s\in [0:1]^N} \!\!\lambda_s^{\star} \left (
\max_{i \in \mathcal{A}_{\rm{S}}} \ell_{i,s}^{\prime} + \max_{i \in \mathcal{N}_{\rm{S}} \backslash \mathcal{A}_{\rm{S}}} r_{i,s}^{\prime}\right )
\\  \geq & \min_{\mathcal{A}_{\rm{F}}\subseteq \mathcal{N}_{\rm{F}} } \left \{ \sum_{s\in [0:1]^N} \!\!\lambda_s^{\star} \left ( \max_{i \in \mathcal{A}_{\rm{F}}} \ell_{i,s}^{\prime} + \max_{i \in \mathcal{N}_{\rm{F}} \backslash \mathcal{A}_{\rm{F}}} r_{i,s}^{\prime}
\right ) \right \}
\\&+ 
\min_{\mathcal{A}_{\rm{S}}\subseteq \mathcal{N}_{\rm{S}} } \left \{
\sum_{s\in [0:1]^N} \!\!\lambda_s^{\star} \left (
\max_{i \in \mathcal{A}_{\rm{S}}} \ell_{i,s}^{\prime}+ \max_{i \in \mathcal{N}_{\rm{S}} \backslash \mathcal{A}_{\rm{S}}} r_{i,s}^{\prime} \right )
\right \}
\\ = & {\mathsf{C}}^{{\rm{HD}}}_{\mathcal{N}_{\rm{F}}} + {\mathsf{C}}^{\star{\rm{HD}}}_{\{1\}},
\end{align*}
where $\ell_{i,s}^{\prime}$ and $r_{i,s}^{\prime}$ are defined in \eqref{eq:NewChannels}
and where the inequality in $\rm{(a)}$ follows from Lemma \ref{prope:FDcut}.
\section{Proof of Theorem \ref{th:theoremDiamFD}}
\label{app:AppFDHD}

\subsection{The case $N=k, \ k \in [1:2]$}
We here prove the result in Theorem~\ref{th:theoremDiamFD} for the case $N=k, \ k \in [1:2]$.
For this case, we trivially have
\begin{align*}
\frac{\mathsf{C}^{{\rm{HD}}}_{k,N}}{{\mathsf{C}}^{{\rm{FD}}}_{\mathcal{N}_{\rm{F}}}}
=
\frac{\mathsf{C}^{{\rm{HD}}}_{k,k}}{{\mathsf{C}}^{{\rm{FD}}}_{\mathcal{N}_{\rm{F}}}}
=
\frac{\mathsf{C}^{{\rm{HD}}}_{k,k}}{2 \tilde{\mathsf{C}}^{{\rm{HD}}}_{k,k}}
\geq
\frac{\tilde{\mathsf{C}}^{{\rm{HD}}}_{k,k}}{2 \tilde{\mathsf{C}}^{{\rm{HD}}}_{k,k}}
=
\frac{1}{2},
\end{align*}
where $\tilde{\mathsf{C}}^{{\rm{HD}}}_{k,k}$ is the HD capacity of the $k$-relay subnetwork (with $k \in [1:2]$) when the relays receive for $\frac{1}{2}$ of the time and transmit for $\frac{1}{2}$ of the time.
We next provide a network example for which $\mathsf{C}^{{\rm{HD}}}_{k,k} = \frac{1}{2} {\mathsf{C}}^{{\rm{FD}}}_{\mathcal{N}_{\rm{F}}}$, with $k \in [1:2]$.
\\
{\bf{Example.}}
For $N=k=1$, consider $\ell_1=r_1=1$; clearly for this network we have ${\mathsf{C}}^{{\rm{FD}}}_{\mathcal{N}_{\rm{F}}} \!=\! 1$ and $\mathsf{C}^{{\rm{HD}}}_{1,1} = \frac{\ell_1 r_1}{\ell_1\!+\!r_1}\!=\!\frac{1}{2}$; hence $\mathsf{C}^{{\rm{HD}}}_{1,1} \!=\!\frac{1}{2}{\mathsf{C}}^{{\rm{FD}}}_{\mathcal{N}_{\rm{F}}}$. 
For $N=k=2$, consider $\ell_1 \!=\! r_1 \!=\! 1, \ell_{2} \!=\! r_{2} \!=\! 0$;
clearly for this network we have ${\mathsf{C}}^{{\rm{FD}}}_{\mathcal{N}_{\rm{F}}} \!=\!1$ and $\mathsf{C}^{{\rm{HD}}}_{2,2} \!=\! \frac{\ell_1 r_1}{\ell_1+r_1}\!=\!\frac{1}{2}$; hence $\mathsf{C}^{{\rm{HD}}}_{2,2} \!=\!\frac{1}{2}{\mathsf{C}}^{{\rm{FD}}}_{\mathcal{N}_{\rm{F}}}$.

\subsection{The case $k=1$ and $N=2$}
We here prove the result in Theorem~\ref{th:theoremDiamFD} for the case $k=1$ and $N=2$.
To this end, we let ${\mathsf{C}}^{{\rm{FD}}}_{\mathcal{N}_{\rm{F}}}= \mathsf{c}$, for some given $\mathsf{c} \in \mathbb{R}_+$ and we assume, without loss of generality, that $\ell_1 \geq \ell_2$. 
As such, proving the lower bound on $\mathsf{C}^{{\rm{HD}}}_{1,2}$ is equivalent to solve the following problem
\begin{align}
\label{eq:2relaysWorstScenario}
\begin{array}{lll}
& f_1:={\rm{min}} & \max_{i \in [1:2]} \left \{\frac{\ell_i r_i}{\ell_i+r_i} \right \}
\\ &{\rm{s.\ t.}} &  \mathsf{c} = \min \left \{ \ell_1, \ell_2+r_1,\max \left \{ r_1,r_2\right \} \right \},
\\ &{\rm{and}} & \ell_1 \geq \ell_2,
\\ &{\rm{and}} &\min \left \{ r_1,\ell_2, r_2 \right \} \geq 0.
\end{array}
\end{align}
Notice that the solution $f_1$ of the problem in \eqref{eq:2relaysWorstScenario} is greater than the solution $f_2$ of the following problem
\begin{align}
\label{eq:2relaysWorstScenarioNew}
\begin{array}{lll}
& f_2:={\rm{min}} & \max_{i \in [1:2]} \left \{\frac{\ell_i r_i}{\ell_i+r_i} \right \}
\\ &{\rm{s.\ t.}} &  \ell_1 \geq \mathsf{c}, \ \ell_2+r_1 \geq \mathsf{c}, \ 
\max \left \{ r_1,r_2\right \} \geq \mathsf{c},
\\ &{\rm{and}} & \ell_1 \geq \ell_2,
\\ &{\rm{and}} &\min \left \{ r_1,\ell_2, r_2 \right \} \geq 0.
\end{array}
\end{align}
The reason why we have $f_1 \geq f_2$ is because we are minimizing the objective function and in the problem in \eqref{eq:2relaysWorstScenarioNew} we are increasing the space of the search with respect to the problem in \eqref{eq:2relaysWorstScenario}.
Thus, if we are able to show that $f_2 \geq \frac{1}{3} \mathsf{c}$, then this implies that also $f_1 \geq \frac{1}{3} \mathsf{c}$.
Hence, we now focus on the problem in \eqref{eq:2relaysWorstScenarioNew}, which can be equivalently rewritten as
\begin{align}
\label{eq:2relaysWorstScenarioNewEquivalent}
\begin{array}{lll}
& {\rm{min}} & t
\\ &{\rm{s.\ t.}} &  \frac{\ell_1 r_1}{\ell_1+r_1} \leq t,
\\ &{\rm{and}} & \frac{\ell_2 r_2}{\ell_2+r_2}  \leq t,
\\ &{\rm{and}} &  \ell_1 \geq \mathsf{c}, \ \ell_2+r_1 \geq \mathsf{c}, \ 
\max \left \{ r_1,r_2\right \} \geq \mathsf{c},
\\ &{\rm{and}} & \ell_1 \geq \ell_2,
\\ &{\rm{and}} &\min \left \{ r_1,\ell_2, r_2,t \right \} \geq 0.
\end{array}
\end{align}
We now analyze two cases:

\noindent $\bullet$ {\bf{Case (i)}}: $r_1 \geq r_2$; 
then, the problem in \eqref{eq:2relaysWorstScenarioNewEquivalent} becomes
\begin{align}
\label{eq:2relaysWorstScenarioNewEquivalentCase1}
\begin{array}{lll}
& {\rm{min}} & \frac{\ell_1 r_1}{\ell_1+r_1}
\\ &{\rm{and}} &  \ell_1 \geq \mathsf{c}, \ \ell_2+r_1 \geq \mathsf{c}, \ 
 r_1 \geq \mathsf{c},
\\ &{\rm{and}} & \ell_1 \geq \ell_2, \ r_1 \geq r_2,
\\ &{\rm{and}} &\min \left \{ \ell_2, r_2 \right \} \geq 0.
\end{array}
\end{align}
Since we have $\ell_1 \geq \mathsf{c}$ and $r_1 \geq \mathsf{c}$ and the objective function in \eqref{eq:2relaysWorstScenarioNewEquivalentCase1} is increasing in $\ell_1$ and $r_1$, we get that the optimal solution is $f_1 \geq \frac{1}{2} \mathsf{c}$, which implies $\mathsf{C}^{{\rm{HD}}}_{1,2} \geq \frac{1}{2}{\mathsf{C}}^{{\rm{FD}}}_{\mathcal{N}_{\rm{F}}}$.

\noindent $\bullet$ {\bf{Case (ii)}}: $r_1 < r_2$; then, the problem in \eqref{eq:2relaysWorstScenarioNewEquivalent} becomes
\begin{align}
\label{eq:2relaysWorstScenarioNewEquivalentCase2}
\begin{array}{lll}
& {\rm{min}} & t
\\ &{\rm{s.\ t.}} &  \frac{\ell_1 r_1}{\ell_1+r_1} \leq t,
\\ &{\rm{and}} & \frac{\ell_2 r_2}{\ell_2+r_2}  \leq t,
\\ &{\rm{and}} &  \ell_1 \geq \mathsf{c}, \ \ell_2+r_1 \geq \mathsf{c}, \ 
r_2\geq \mathsf{c},
\\ &{\rm{and}} & \ell_1 \geq \ell_2, \ r_1 < r_2,
\\ &{\rm{and}} &\min \left \{ r_1,\ell_2,t \right \} \geq 0.
\end{array}
\end{align}
It is not difficult to see that if $r_1>\mathsf{c}$, then the optimal solution would be $f_1 > \frac{1}{2} \mathsf{c}$ since $\ell_1 \geq \mathsf{c}$, which implies $\mathsf{C}^{{\rm{HD}}}_{1,2} > \frac{1}{2}{\mathsf{C}}^{{\rm{FD}}}_{\mathcal{N}_{\rm{F}}}$.
Thus, we now focus on the case $r_1 \leq \mathsf{c}$.
For this case, we can set $\ell_2 = \mathsf{c}-r_1$, without loss of optimality (since $\frac{\ell_2 r_2}{\ell_2+r_2}$ is an increasing function in $\ell_2$); with this we obtain
\begin{align}
\label{eq:2relaysWorstScenarioNewEquivalentCase2a}
\begin{array}{lll}
& {\rm{min}} & t
\\ &{\rm{s.\ t.}} &  \frac{\ell_1 r_1}{\ell_1+r_1} \leq t,
\\ &{\rm{and}} & \frac{ \left( \mathsf{c}-r_1\right) r_2}{\mathsf{c}-r_1+r_2}  \leq t,
\\ &{\rm{and}} &  \ell_1 \geq \mathsf{c}, \ 
r_2\geq \mathsf{c}, \ r_1 < r_2, \ r_1 \leq \mathsf{c}
\\ &{\rm{and}} &\min \left \{ r_1,t \right \} \geq 0.
\end{array}
\end{align}
Now, in the problem in \eqref{eq:2relaysWorstScenarioNewEquivalentCase2a} we can set $\ell_1 = \mathsf{c}$, without loss of optimality (since $\frac{\ell_1r_1}{\ell_1+r_1}$ is an increasing function in $\ell_1$); with this we obtain
\begin{align}
\label{eq:2relaysWorstScenarioNewEquivalentCase2b}
\begin{array}{lll}
& {\rm{min}} & t
\\ &{\rm{s.\ t.}} &  \frac{\mathsf{c} r_1}{\mathsf{c}+r_1} \leq t,
\\ &{\rm{and}} & \frac{ \left( \mathsf{c}-r_1\right) r_2}{\mathsf{c}-r_1+r_2}  \leq t,
\\ &{\rm{and}} &  
r_2\geq \mathsf{c}, \ r_1 < r_2, \ r_1 \leq \mathsf{c}
\\ &{\rm{and}} &\min \left \{ r_1,t \right \} \geq 0.
\end{array}
\end{align}
Now, in the problem in \eqref{eq:2relaysWorstScenarioNewEquivalentCase2b} we can set $r_2 = \mathsf{c}$ (since $\frac{\ell_2 r_2}{\ell_2+r_2}$ is an increasing function in $r_2$), without loss of optimality; with this we obtain
\begin{align}
\label{eq:2relaysWorstScenarioNewEquivalentCase2c}
\begin{array}{lll}
& {\rm{min}} & t
\\ &{\rm{s.\ t.}} &  \frac{\mathsf{c} r_1}{\mathsf{c}+r_1} \leq t,
\\ &{\rm{and}} & \frac{ \left( \mathsf{c}-r_1\right) \mathsf{c}}{2\mathsf{c}-r_1}  \leq t,
\\ &{\rm{and}} &  0 \leq r_1 < \mathsf{c}.
\end{array}
\end{align}
It is not difficult to see that if $r_1 \geq \frac{\mathsf{c}}{2}$, then $\frac{\mathsf{c} r_1}{\mathsf{c}+r_1} \geq \frac{ \left( \mathsf{c}-r_1\right) \mathsf{c}}{2\mathsf{c}-r_1}$, leading to
\begin{align}
\label{eq:2relaysWorstScenarioNewEquivalentCase2d}
\begin{array}{lll}
& {\rm{min}} & \frac{\mathsf{c} r_1}{\mathsf{c}+r_1}
\\ &{\rm{and}} &  \frac{\mathsf{c}}{2} \leq r_1 < \mathsf{c},
\end{array}
\end{align}
which has $f_1 = \frac{1}{3} \mathsf{c}$ as optimal solution. 
This implies $\mathsf{C}^{{\rm{HD}}}_{1,2} = \frac{1}{3}{\mathsf{C}}^{{\rm{FD}}}_{\mathcal{N}_{\rm{F}}}$.
Similarly, if $r_1 \leq \frac{\mathsf{c}}{2}$, then $\frac{\mathsf{c} r_1}{\mathsf{c}+r_1} \leq \frac{ \left( \mathsf{c}-r_1\right) \mathsf{c}}{2\mathsf{c}-r_1}$, leading to
\begin{align}
\label{eq:2relaysWorstScenarioNewEquivalentCase2e}
\begin{array}{lll}
& {\rm{min}} & \frac{ \left( \mathsf{c}-r_1\right) \mathsf{c}}{2\mathsf{c}-r_1}
\\ &{\rm{and}} &  0 \leq r_1 \leq  \frac{1}{2}\mathsf{c},
\end{array}
\end{align}
which has $f_1 = \frac{1}{3} \mathsf{c}$ as optimal solution. 
This implies $\mathsf{C}^{{\rm{HD}}}_{1,2} = \frac{1}{3}{\mathsf{C}}^{{\rm{FD}}}_{\mathcal{N}_{\rm{F}}}$.
This concludes the proof that $\mathsf{C}^{{\rm{HD}}}_{1,2} \geq \frac{1}{3}{\mathsf{C}}^{{\rm{FD}}}_{\mathcal{N}_{\rm{F}}}$.
We next provide a network example for which $\mathsf{C}^{{\rm{HD}}}_{1,2} = \frac{1}{3} {\mathsf{C}}^{{\rm{FD}}}_{\mathcal{N}_{\rm{F}}}$.
\\
{\bf{Example.}}
Consider
$\ell_1 = 1, \ell_{2} = r_{1} = \frac{1}{2}$ and $r_2=1$;
clearly for this network we have ${\mathsf{C}}^{{\rm{FD}}}_{\mathcal{N}_{\rm{F}}} =1$ and $\mathsf{C}^{{\rm{HD}}}_{1,2} = \frac{\ell_1 r_1}{\ell_1+r_1}=\frac{1}{3}$; hence $\mathsf{C}^{{\rm{HD}}}_{1,2} =\frac{1}{3}{\mathsf{C}}^{{\rm{FD}}}_{\mathcal{N}_{\rm{F}}}$.

\subsection{The case $k=1$ and $N \gg 1$}
\begin{figure}
\begin{center}
\includegraphics[width=\columnwidth]{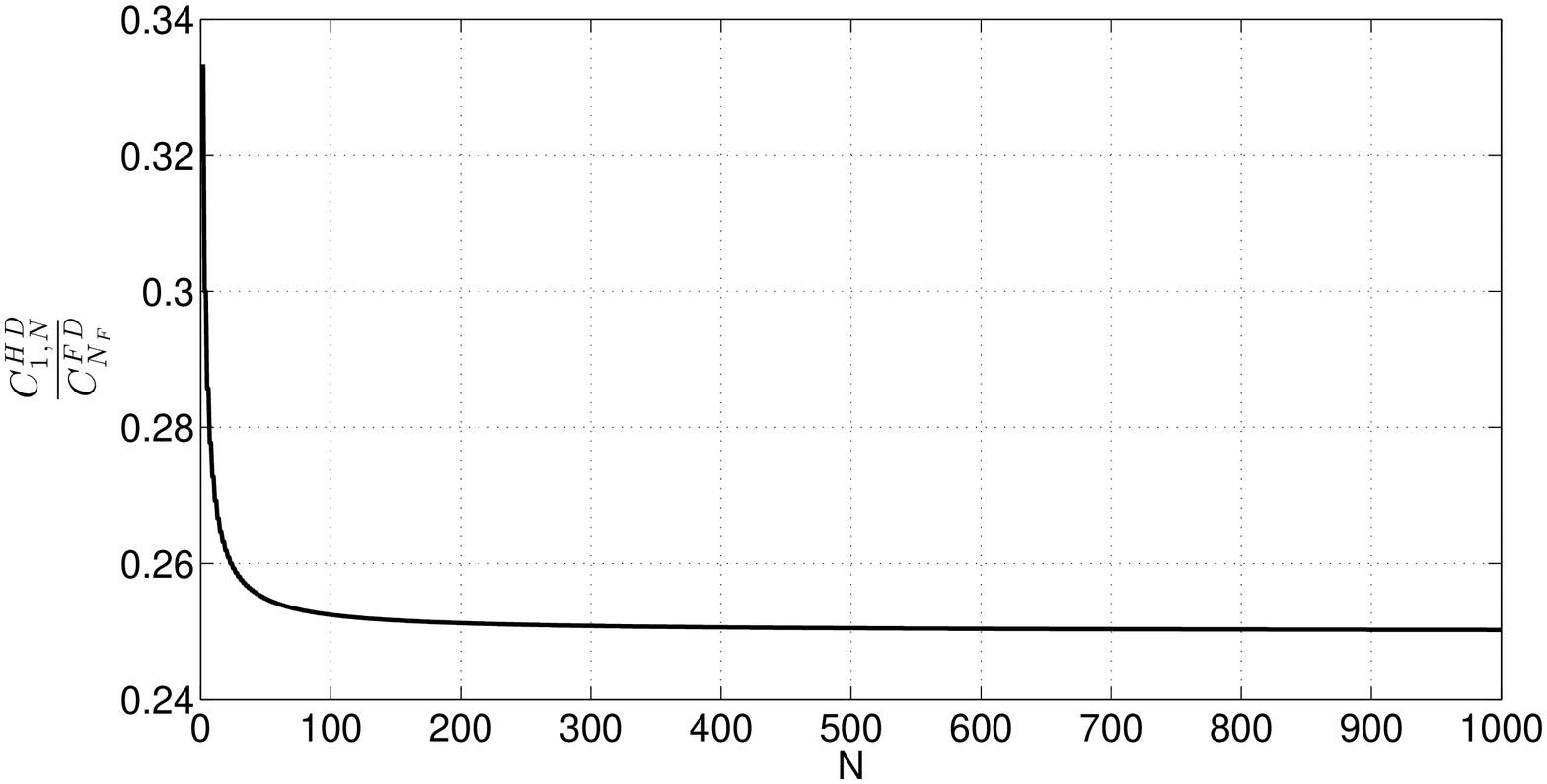}
\end{center}
\vspace{-6mm}
\caption{$\frac{\mathsf{C}^{{\rm{HD}}}_{1,N}}{{\mathsf{C}}^{{\rm{FD}}}_{\mathcal{N}_{\rm{F}}}}$ for the network in \eqref{eq:worstrelayHDex} with $\mathsf{c}=1$.}
\label{fig:fractionkrelayFDupper1}
\vspace{-3mm}
\end{figure}

We here prove the result in Theorem~\ref{th:theoremDiamFD} for the case $k=1$ and $N \gg 1$.
To this end we make use of the result derived in~\cite[Theorem 1]{NazarogluIT2014} for this case, namely
\begin{align*}
\mathsf{C}^{{\rm{FD}}}_{1} \geq 
\frac{1}{2}{\mathsf{C}}^{{\rm{FD}}}_{\mathcal{N}_{\rm{F}}},
\end{align*} 
where we used the notation $\mathsf{C}^{{\rm{FD}}}_{1}$ (which indicates the FD capacity of the $k=1$ selected relay) to highlight that in FD the ratio $\frac{\mathsf{C}^{{\rm{FD}}}_{1}}{{\mathsf{C}}^{{\rm{FD}}}_{\mathcal{N}_{\rm{F}}}}$ does not depend on $N$.
It is not difficult to see that $\mathsf{C}^{{\rm{FD}}}_{1} = 2 \tilde{\mathsf{C}}^{{\rm{HD}}}_{1,N}$, where $\tilde{\mathsf{C}}^{{\rm{HD}}}_{1,N}$ is the HD capacity of the $k=1$ selected relay when it receives for $\frac{1}{2}$ of the time and it transmits for $\frac{1}{2}$ of the time. Thus,
\begin{align*}
\frac{1}{2}{\mathsf{C}}^{{\rm{FD}}}_{\mathcal{N}_{\rm{F}}} \leq
\mathsf{C}^{{\rm{FD}}}_{1}=
2 \tilde{\mathsf{C}}^{{\rm{HD}}}_{1,N} \leq
2\mathsf{C}^{{\rm{HD}}}_{1,N}
\Longrightarrow
\frac{\mathsf{C}^{{\rm{HD}}}_{1,N}}{{\mathsf{C}}^{{\rm{FD}}}_{\mathcal{N}_{\rm{F}}}}
\geq \frac{1}{4}.
\end{align*}
We next provide a network example for which $\mathsf{C}^{{\rm{HD}}}_{1,N}= \frac{1}{4} {\mathsf{C}}^{{\rm{FD}}}_{\mathcal{N}_{\rm{F}}}$ for $N \gg 1$.
\\
{\bf{Example.}} We consider the network defined in \eqref{eq:worstrelayHDex}
for which all the single-relay HD capacities are the same and given in~\eqref{eq:singlecapHD} and ${\mathsf{C}}^{{\rm{FD}}}_{\mathcal{N}_{\rm{F}}} = \mathsf{c}$.
With this we obtain the claimed ratio $\mathsf{C}^{{\rm{HD}}}_{1,N}= \frac{1}{4} {\mathsf{C}}^{{\rm{FD}}}_{\mathcal{N}_{\rm{F}}}$ for $N \gg 1$ as we also observe from Fig.~\ref{fig:fractionkrelayFDupper1}.

\subsection{The case $k=2$ and $N=3$}
We here prove the result in Theorem~\ref{th:theoremDiamFD} for the case $k=2$ and $N \gg 1$. To this end we make use of the following upper bound
\begin{align*}
{\mathsf{C}}^{{\rm{HD}}}_{\mathcal{N}_{\rm{F}}} \leq {\mathsf{C}}^{{\rm{FD}}}_{\mathcal{N}_{\rm{F}}},
\end{align*}
and we let $\bar{\mathsf{C}}^{{\rm{HD}}}_{jk}$ be the HD capacity of the subnetwork of relays $j$ and $k$ with $\left( j,k\right) \in [1:N]^2$ and $j \neq k$.

Without loss of generality, we let ${\mathsf{C}}^{{\rm{FD}}}_{\mathcal{N}_{\rm{F}}} = \mathsf{c}$ and we assume $\ell_1 \geq \ell_2 \geq \ell_3$.
With these assumptions we need to have
\begin{align}
\label{eq:cBefore}
\mathsf{c}& = \min \left \{ \ell_1, \ell_2 + r_1,
 \ell_3 + \max \left \{r_1,r_2 \right \}, \right. \nonumber
\\& \left. \qquad \max \left \{r_1,r_2,r_3 \right \} \right\}.
\end{align}
Since we are interested in finding the worst possible ratio, the above constraint can be relaxed as
\begin{subequations}
\label{eq:cAfter}
\begin{align}
&\ell_1 \geq \mathsf{c},
\\& \ell_2 + r_1 \geq \mathsf{c},
\\ & \ell_3 + \max \left \{r_1,r_2 \right \} \geq \mathsf{c},
\\ & \max \left \{r_1,r_2,r_3 \right \} \geq \mathsf{c}.
\end{align}
\end{subequations}
The reason why we can relax the constraint in \eqref{eq:cBefore} as those in \eqref{eq:cAfter} is because in \eqref{eq:cAfter} we are enlarging the space of the search and because we are interested in the worst possible ratio.
It is not difficult to see that, since the capacity is a non-decreasing function of the channel parameters, we can set, without loss of generality, $\ell_1 = \mathsf{c}$.
With this, we need to have
\begin{subequations}
\label{eq:CutConstr}
\begin{align}
& \ell_2 + r_1 \geq \mathsf{c},
\label{eq:CutConstr1}
\\ & \ell_3 + \max \left \{r_1,r_2 \right \} \geq \mathsf{c},
\label{eq:CutConstr2}
\\ & \max \left \{r_1,r_2,r_3 \right \} \geq \mathsf{c}.
\label{eq:CutConstr3}
\end{align}
\end{subequations}
In what follows we will show by contradiction that at least one of the three possible subnetworks of $k=2$ relays has a HD capacity greater than or equal to $\frac{1}{2} {\mathsf{C}}^{{\rm{FD}}}_{\mathcal{N}_{\rm{F}}}$. 
\\
{\bf{Case (i):}}
$r_1 r_2 \geq \mathsf{c} \ell_2$; with this we obtain that the capacity $\bar{\mathsf{C}}^{{\rm{HD}}}_{12}$ of the first pair (i.e., relay 1 and relay 2) is given by \cite{BagheriIT2014}
\begin{align*}
\bar{\mathsf{C}}^{{\rm{HD}}}_{12} = \frac{\mathsf{c} \ell_2 r_1 - \ell_2^2 r_2 + \mathsf{c} \ell_2 r_2 + \mathsf{c} r_1 r_2}
{\left( \ell_2 + r_2\right) \left(\mathsf{c}-\ell_2+r_1 \right)},
\end{align*}
which implies $\frac{\bar{\mathsf{C}}^{{\rm{HD}}}_{12}}{{\mathsf{C}}^{{\rm{FD}}}_{\mathcal{N}_{\rm{F}}}} = \frac{\mathsf{c} \ell_2 r_1 - \ell_2^2 r_2 + \mathsf{c} \ell_2 r_2 + \mathsf{c} r_1 r_2}
{\mathsf{c}\left( \ell_2 + r_2\right) \left(\mathsf{c}-\ell_2+r_1 \right)}$. We now want to show that, in order to meet the conditions in \eqref{eq:CutConstr}, there are no networks for which
\begin{align*}
\frac{\bar{\mathsf{C}}^{{\rm{HD}}}_{12}}{{\mathsf{C}}^{{\rm{FD}}}_{\mathcal{N}_{\rm{F}}}} = \frac{\mathsf{c} \ell_2 r_1 - \ell_2^2 r_2 + \mathsf{c} \ell_2 r_2 + \mathsf{c} r_1 r_2}
{\mathsf{c}\left( \ell_2 + r_2\right) \left(\mathsf{c}-\ell_2+r_1 \right)}< \frac{1}{2},
\end{align*}
which is equivalent to show that there are no networks for which
\begin{align*}
&\mathsf{c} \ell_2 r_1 - 2 \ell_2^2 r_2 + 3 \mathsf{c} \ell_2 r_2 +\mathsf{c} r_1 r_2 - \ell_2 \mathsf{c}^2 + \mathsf{c} \ell_2^2 - \mathsf{c}^2 r_2 < 0,
\\ \Rightarrow & \mathsf{c} \left( \ell_2+r_2\right) \underbrace{\left( r_1 + \ell_2 - \mathsf{c} \right)}_{\geq 0 \ \text{from} \ \eqref{eq:CutConstr1}}
+ 2 \ell_2 r_2 \underbrace{\left( \mathsf{c}-\ell_2\right)}_{\geq 0 \ \text{since} \ \mathsf{c}=\ell_1 \geq \ell_2} < 0.
\end{align*}
Thus, there are no channel conditions for which $\bar{\mathsf{C}}^{{\rm{HD}}}_{12} <\frac{1}{2} \mathsf{c}$, i.e., by letting relay 1 and relay 2 operate we always achieve (to within a constant gap) at least half of the capacity of the corresponding FD network.
\\
{\bf{Case (ii):}}
$r_1 r_2 \leq \mathsf{c} \ell_2$ and $r_1 \geq r_2$; with this we obtain that the capacity $\bar{\mathsf{C}}^{{\rm{HD}}}_{12}$ of the first pair (i.e., relay 1 and relay 2) is given by \cite{BagheriIT2014}, i.e.,
\begin{align*}
\bar{\mathsf{C}}^{{\rm{HD}}}_{12} = \frac{\mathsf{c} \ell_2 r_1 - \ell_2 r_2^2 + \mathsf{c} r_1 r_2 + \ell_2 r_1 r_2}
{\left( \ell_2 + r_2\right) \left(\mathsf{c}+r_1-r_2 \right)},
\end{align*}
which implies 
$\frac{\bar{\mathsf{C}}^{{\rm{HD}}}_{12}}{{\mathsf{C}}^{{\rm{FD}}}_{\mathcal{N}_{\rm{F}}}} = \frac{\mathsf{c} \ell_2 r_1 - \ell_2 r_2^2 + \mathsf{c} r_1 r_2 + \ell_2 r_1 r_2}
{\mathsf{c}\left( \ell_2 + r_2\right) \left(\mathsf{c}+r_1-r_2 \right)}$.
We now want to show that, in order to meet the conditions in \eqref{eq:CutConstr}, there are no networks for which
\begin{align*}
&\frac{\bar{\mathsf{C}}^{{\rm{HD}}}_{12}}{{\mathsf{C}}^{{\rm{FD}}}_{\mathcal{N}_{\rm{F}}}}= \frac{\mathsf{c} \ell_2 r_1 - \ell_2 r_2^2 + \mathsf{c} r_1 r_2 + \ell_2 r_1 r_2}
{\mathsf{c}\left( \ell_2 + r_2\right) \left(\mathsf{c}+r_1-r_2 \right)} < \frac{1}{2},
\end{align*}
which is equivalent to show that there are no networks for which
\begin{align*}
&\mathsf{c} \ell_2 r_1 - 2 \ell_2 r_2^2 + \mathsf{c}r_1r_2 + 2 \ell_2 r_1 r_2 - \ell_2 \mathsf{c}^2 
\\&+ \ell_2 r_2 \mathsf{c} - r_2 \mathsf{c}^2 + r_2^2 \mathsf{c} < 0
\\ \Rightarrow & 
2\ell_2 r_2 \underbrace{\left( r_1-r_2 \right)}_{\geq 0 \ \text{since} \ r_1 \geq r_2}
+ \mathsf{c} r_2 \underbrace{\left( r_1 + \ell_2 - \mathsf{c}\right)}_{\geq 0 \ \text{from} \ \eqref{eq:CutConstr1}}
\\&+ \mathsf{c} \underbrace{\left( \ell_2 r_1-\ell_2 \mathsf{c} +r_2^2\right)}_{\geq 0 \ \text{if} \ r_1 \geq \mathsf{c}} < 0.
\end{align*}
Thus, a sufficient condition for the above quantity to be always positive is $r_1 \geq \mathsf{c}$, i.e., under these channel conditions we never have $\bar{\mathsf{C}}^{{\rm{HD}}}_{12} <\frac{1}{2} \mathsf{c}$.
We now analyze the case $r_1 < \mathsf{c}$, which, because of the constraint in \eqref{eq:CutConstr3}, implies $r_3 \!\geq\! r_1 \!\geq\! r_2$.
\\
{\bf{Case (ii-a):}} $\mathsf{c} \ell_3 \leq r_1 r_3$; with this we obtain that the capacity $\bar{\mathsf{C}}^{{\rm{HD}}}_{13}$ of the pair relay 1 and relay 3 is given by \cite{BagheriIT2014}, i.e., 
\begin{align*}
\bar{\mathsf{C}}^{{\rm{HD}}}_{13}  = \frac{\mathsf{c} \ell_3 r_1 - \ell_3^2 r_3 + \mathsf{c} \ell_3 r_3 + \mathsf{c} r_1 r_3}
{\left( \ell_3 + r_3\right) \left(\mathsf{c}-\ell_3+r_1 \right)},
\end{align*}
which implies 
$\frac{\bar{\mathsf{C}}^{{\rm{HD}}}_{13}}{{\mathsf{C}}^{{\rm{FD}}}_{\mathcal{N}_{\rm{F}}}}=\frac{\mathsf{c} \ell_3 r_1 - \ell_3^2 r_3 + \mathsf{c} \ell_3 r_3 + \mathsf{c} r_1 r_3}
{\mathsf{c} \left( \ell_3 + r_3\right) \left(\mathsf{c}-\ell_3+r_1 \right)}$.
We now want to show that, in order to meet the conditions in \eqref{eq:CutConstr}, there are no networks for which
\begin{align*}
&\frac{\bar{\mathsf{C}}^{{\rm{HD}}}_{13}}{{\mathsf{C}}^{{\rm{FD}}}_{\mathcal{N}_{\rm{F}}}}= \frac{\mathsf{c} \ell_3 r_1 - \ell_3^2 r_3 + \mathsf{c} \ell_3 r_3 + \mathsf{c} r_1 r_3}
{\mathsf{c} \left( \ell_3 + r_3\right) \left(\mathsf{c}-\ell_3+r_1 \right)} < \frac{1}{2},
\end{align*}
which is equivalent to show that there are no networks for which
\begin{align*}
&\mathsf{c} \ell_3 r_1 - 2 \ell_3^2 r_3 + 3 \mathsf{c} \ell_3 r_3 +\mathsf{c} r_1 r_3 - \ell_3 \mathsf{c}^2 + \mathsf{c} \ell_3^2 - \mathsf{c}^2 r_3 \!<\! 0
\\ \Rightarrow &   \mathsf{c} \left(\ell_3 +r_3 \right) \underbrace{\left( r_1 + \ell_3 - \mathsf{c} \right)}_{\geq 0 \ \text{from} \ \eqref{eq:CutConstr2}}
+ 2 \ell_3 r_3 \underbrace{\left( \mathsf{c}-\ell_3\right)}_{\geq 0 \ \text{since} \ \mathsf{c}=\ell_1 \geq \ell_3} < 0.
\end{align*}
Thus, there are no channel conditions for which $\bar{\mathsf{C}}^{{\rm{HD}}}_{13} <\frac{1}{2} \mathsf{c}$, i.e., by letting relay 1 and relay 3 operate we always achieve (to within a constant gap) at least half of the capacity of the corresponding FD network.
\\
{\bf{Case (ii-b):}} $\mathsf{c} \ell_3 \geq r_1 r_3$; with this we obtain that the capacity $\bar{\mathsf{C}}^{{\rm{HD}}}_{13}$ of the pair relay 1 and relay 3 is given by \cite{BagheriIT2014}, i.e.,
\begin{align}
\bar{\mathsf{C}}^{{\rm{HD}}}_{13}  = \frac{\mathsf{c} \ell_3 r_3 - \mathsf{c} r_1^2 + \mathsf{c} r_1 r_3 + \ell_3 r_1 r_3}
{\left( \mathsf{c} + r_1\right) \left(\ell_3+r_3-r_1 \right)},
\end{align}
which implies 
$\frac{\bar{\mathsf{C}}^{{\rm{HD}}}_{13}}{{\mathsf{C}}^{{\rm{FD}}}_{\mathcal{N}_{\rm{F}}}} = \frac{\mathsf{c} \ell_3 r_3 - \mathsf{c} r_1^2 + \mathsf{c} r_1 r_3 + \ell_3 r_1 r_3}
{\mathsf{c} \left( \mathsf{c} + r_1\right) \left(\ell_3+r_3-r_1 \right)}$.
We now want to show that, in order to meet the conditions in \eqref{eq:CutConstr}, there are no networks for which
\begin{align}
&\frac{\bar{\mathsf{C}}^{{\rm{HD}}}_{13}}{{\mathsf{C}}^{{\rm{FD}}}_{\mathcal{N}_{\rm{F}}}}  = \frac{\mathsf{c} \ell_3 r_3 - \mathsf{c} r_1^2 + \mathsf{c} r_1 r_3 + \ell_3 r_1 r_3}
{\mathsf{c} \left( \mathsf{c} + r_1\right) \left(\ell_3+r_3-r_1 \right)} < \frac{1}{2},
\end{align}
which is equivalent to show that there are no networks for which
\begin{align*}
&2 \mathsf{c} \ell_3 r_3 - \mathsf{c} r_1^2 + \mathsf{c} r_1 r_3 + 2 \ell_3 r_1 r_3 
\\&- \mathsf{c}^2 \ell_3 - \mathsf{c}^2 r_3 + \mathsf{c}^2 r_1 - \mathsf{c} r_1 \ell_3 < 0
\\ \Rightarrow &
 \ell_3 \left(\mathsf{c} +r_1\right) \underbrace{\left( r_3 - \mathsf{c} \right)}_{\geq 0 \ \text{from} \ \eqref{eq:CutConstr3}}
+ \mathsf{c} r_3 \underbrace{\left( \ell_3 + r_1 - \mathsf{c}\right)}_{\geq 0 \ \text{from} \ \eqref{eq:CutConstr2}}
\\&+ \mathsf{c} r_1 \underbrace{\left( \mathsf{c} - r_1 \right)}_{\geq 0 \ \text{since} \ r_1 < \mathsf{c}}
+ \ell_3 r_1 r_3 <0.
\end{align*}
Thus, there are no channel conditions for which $\bar{\mathsf{C}}^{{\rm{HD}}}_{13} <\frac{1}{2} \mathsf{c}$, i.e., by letting relay 1 and relay 3 operate we always achieve (to within a constant gap) at least half of the capacity of the corresponding FD network.
\\
{\bf{Case (iii):}}
$r_1 r_2 \leq \mathsf{c} \ell_2$ and $r_1 \leq r_2$; 
with this we obtain that the capacity $\bar{\mathsf{C}}^{{\rm{HD}}}_{12}$ of the first pair (i.e., relay 1 and relay 2) is given by \cite{BagheriIT2014}, i.e.,
\begin{align}
\bar{\mathsf{C}}^{{\rm{HD}}}_{12} =\frac{\mathsf{c} \ell_2 r_2 - \mathsf{c} r_1^2 + \mathsf{c} r_1 r_2 + \ell_2 r_1 r_2}
{\left( \mathsf{c} + r_1\right) \left(\ell_2+r_2-r_1 \right)},
\end{align}
which implies $\frac{\bar{\mathsf{C}}^{{\rm{HD}}}_{12}}{{\mathsf{C}}^{{\rm{FD}}}_{\mathcal{N}_{\rm{F}}}} =\frac{\mathsf{c} \ell_2 r_2 - \mathsf{c} r_1^2 + \mathsf{c} r_1 r_2 + \ell_2 r_1 r_2}
{\mathsf{c}\left( \mathsf{c} + r_1\right) \left(\ell_2+r_2-r_1 \right)}$.
We now want to show that, in order to meet the conditions in \eqref{eq:CutConstr}, there are no networks for which
\begin{align*}
&\frac{\bar{\mathsf{C}}^{{\rm{HD}}}_{12}}{{\mathsf{C}}^{{\rm{FD}}}_{\mathcal{N}_{\rm{F}}}} = \frac{\mathsf{c} \ell_2 r_2 - \mathsf{c} r_1^2 + \mathsf{c} r_1 r_2 + \ell_2 r_1 r_2}
{\mathsf{c}\left( \mathsf{c} + r_1\right) \left(\ell_2+r_2-r_1 \right)} < \frac{1}{2},
\end{align*}
which is equivalent to show that there are no networks for which
\begin{align*}
&2 \mathsf{c} \ell_2 r_2 - \mathsf{c} r_1^2 + \mathsf{c} r_1 r_2 + 2 \ell_2 r_1 r_2 
\\& -\mathsf{c}^2 \ell_2 - \mathsf{c}^2 r_2 + \mathsf{c}^2 r_1 - \mathsf{c} r_1 \ell_2 < 0
\\ \Rightarrow &
\mathsf{c} \underbrace{\left( r_2-r_1\right) }_{\geq 0 \ \text{since} \ r_2 \geq r_1}
\underbrace{\left( \ell_2 + r_1 - \mathsf{c}\right)}_{\geq 0 \ \text{from} \ \eqref{eq:CutConstr1}}
+  \ell_2 \underbrace{\left( r_2 \mathsf{c} + 2  r_1 r_2 - \mathsf{c}^2 \right)}_{\geq 0 \ \text{if} \ r_2 \geq \frac{\mathsf{c}^2}{\mathsf{c}+2r_1}}
 < 0.
\end{align*}
Thus, a sufficient condition for the above quantity to be always positive is $r_2 \geq \frac{\mathsf{c}^2}{\mathsf{c}+2r_1}$, i.e., under these channel conditions we never have $\bar{\mathsf{C}}^{{\rm{HD}}}_{12}<\frac{1}{2} \mathsf{c}$. 
Thus, we now analyze the case $r_2 < \frac{\mathsf{c}^2}{\mathsf{c}+2r_1}$, which, because of the constraint in \eqref{eq:CutConstr3}, implies $r_3 \geq r_2 \geq r_1$.
\\
{\bf{Case (iii-a):}} $\ell_2 \ell_3 \leq r_2 r_3$; 
with this we obtain that the capacity $\bar{\mathsf{C}}^{{\rm{HD}}}_{23}$ of the pair relay 2 and relay 3 is given by \cite{BagheriIT2014}, i.e., 
\begin{align}
\bar{\mathsf{C}}^{{\rm{HD}}}_{23} = \frac{\ell_2 \ell_3 r_2 - \ell_3^2 r_3 + \ell_2 \ell_3 r_3 + \ell_2 r_2 r_3}
{\left( \ell_3 + r_3\right) \left(\ell_2-\ell_3+r_2 \right)},
\end{align}
which implies 
$\frac{\bar{\mathsf{C}}^{{\rm{HD}}}_{23}}{{\mathsf{C}}^{{\rm{FD}}}_{\mathcal{N}_{\rm{F}}}} =\frac{\ell_2 \ell_3 r_2 - \ell_3^2 r_3 + \ell_2 \ell_3 r_3 + \ell_2 r_2 r_3}
{\mathsf{c} \left( \ell_3 + r_3\right) \left(\ell_2-\ell_3+r_2 \right)}$.
We now want to show that, in order to meet the conditions in \eqref{eq:CutConstr}, there are no networks for which
\begin{align*}
&\frac{\bar{\mathsf{C}}^{{\rm{HD}}}_{23}}{{\mathsf{C}}^{{\rm{FD}}}_{\mathcal{N}_{\rm{F}}}}  = \frac{\ell_2 \ell_3 r_2 - \ell_3^2 r_3 + \ell_2 \ell_3 r_3 + \ell_2 r_2 r_3}
{\mathsf{c} \left( \ell_3 + r_3\right) \left(\ell_2-\ell_3+r_2 \right)} < \frac{1}{2},
\end{align*}
which is equivalent to show that there are no networks for which
\begin{align}
\label{eq:finalmente1}
&2 \ell_2 \ell_3 r_2 - 2 \ell_3^2 r_3 + 2 \ell_2 \ell_3 r_3 + 2 \ell_2 r_2 r_3 - \mathsf{c} \ell_2 \ell_3 + \mathsf{c} \ell_3^2 \nonumber
\\&- \mathsf{c} \ell_3 r_2 - \mathsf{c} r_3 \ell_2 + \mathsf{c} r_3 \ell_3 - \mathsf{c} r_2 r_3 < 0.
\end{align}
It is not difficult to see that the Left-Hand Side (LHS) of \eqref{eq:finalmente1} is always increasing in $\ell_2$; the derivative of the LHS of \eqref{eq:finalmente1} with respect to $\ell_2$ is, in fact, given by
\begin{align*}
&2 \ell_3 r_2 + 2 \ell_3 r_3 + 2 r_2 r_3 - \mathsf{c} \ell_3 - \mathsf{c} r_3
\\&= 2 \ell_3 r_2 + r_2 r_3 
+ \ell_3 \underbrace{\left( r_3 - \mathsf{c} \right)}_{\geq 0 \ \text{from} \ \eqref{eq:CutConstr3}}
+ r_3 \underbrace{\left( \ell_3 + r_2 - \mathsf{c} \right)}_{\geq 0 \ \text{from} \ \eqref{eq:CutConstr2}} \geq 0.
\end{align*}
Hence, from \eqref{eq:finalmente1} we obtain
\begin{align*}
&2 \ell_2 \ell_3 r_2 - 2 \ell_3^2 r_3 + 2 \ell_2 \ell_3 r_3 + 2 \ell_2 r_2 r_3 - \mathsf{c} \ell_2 \ell_3 + \mathsf{c} \ell_3^2 
\\& - \mathsf{c} \ell_3 r_2 - \mathsf{c} r_3 \ell_2 + \mathsf{c} r_3 \ell_3 - \mathsf{c} r_2 r_3
\\ & \stackrel{\ell_2 \geq \ell_3}{ \geq } 2 \ell_3^2 r_2 + 2 \ell_3 r_2 r_3 - \mathsf{c} \ell_3 r_2 - \mathsf{c}r_2 r_3
\\&= r_2\left( r_3 + \ell_3 \right)
\underbrace{\left( 2 \ell_3 - \mathsf{c} \right)}_{\geq 0 \ \text{if} \ \ell_3 \geq \frac{\mathsf{c}}{2}}.
\end{align*}
Thus, a sufficient condition for the above quantity to be always positive is $\ell_3 \geq \frac{\mathsf{c}}{2}$, i.e., under these channel conditions we never have $\bar{\mathsf{C}}^{{\rm{HD}}}_{23} < \frac{1}{2} \mathsf{c}$. 
Thus we now analyze the case $\ell_3 < \frac{\mathsf{c}}{2}$. We start by noticing that the LHS of \eqref{eq:finalmente1} is always increasing in $r_2$; the derivative of the LHS of \eqref{eq:finalmente1} with respect to $r_2$ is in fact given by
\begin{align*}
&2 \ell_2 \ell_3 + 2 \ell_2 r_3 - \mathsf{c} \ell_3 - \mathsf{c} r_3 = \left(\ell_3+r_3 \right) 
\left(2 \ell_2 - \mathsf{c} \right) \geq 0.
\end{align*}
Notice that the fact that $\ell_2 \geq \frac{\mathsf{c}}{2}$ follows since:
(i) from \eqref{eq:CutConstr2} $\ell_3 < \frac{\mathsf{c}}{2}$ implies $r_2 > \frac{\mathsf{c}}{2}$ 
and (ii)
since $r_1 \leq r_2 < \frac{\mathsf{c}^2}{\mathsf{c}+2r_1}$ implies $r_1 < \frac{\mathsf{c}}{2}$ from \eqref{eq:CutConstr1} we must have $\ell_2 > \frac{\mathsf{c}}{2}$. 
Hence, from \eqref{eq:finalmente1} we obtain
\begin{align*}
&2 \ell_2 \ell_3 r_2 - 2 \ell_3^2 r_3 + 2 \ell_2 \ell_3 r_3 + 2 \ell_2 r_2 r_3 - \mathsf{c} \ell_2 \ell_3 + \mathsf{c} \ell_3^2 
\\& \quad - \mathsf{c} \ell_3 r_2 - \mathsf{c} r_3 \ell_2 + \mathsf{c} r_3 \ell_3 - \mathsf{c} r_2 r_3
\\& \stackrel{r_2 \geq \mathsf{c}-\ell_3}{\geq}
\ell_2 \ell_3 \mathsf{c} - 2 \ell_2 \ell_3^2 - 2 \ell_3^2 r_3 + \ell_2 r_3 \mathsf{c} + 2 \mathsf{c} \ell_3^2 
\\& \quad -\mathsf{c}^2 \ell_3 +2\mathsf{c} r_3 \ell_3 - \mathsf{c}^2 r_3
\\& = \ell_3 
\underbrace{\left( \mathsf{c}-2\ell_3 \right)}_{> 0 \ \text{since} \ \ell_3 < \frac{\mathsf{c}}{2}}
 \underbrace{\left( r_3 - \mathsf{c}\right)}_{\geq 0 \ \text{since} \ r_3 \geq \mathsf{c}}
+ \ell_2 \ell_3 
\underbrace{\left( \mathsf{c}-2 \ell_3 \right)}_{> 0 \ \text{since} \ \ell_3 < \frac{\mathsf{c}}{2}}
\\& \quad + \mathsf{c} r_3 \left( \ell_2+\ell_3-\mathsf{c}\right).
\end{align*}
We next show that the term $\left(\ell_2 + \ell_3 - \mathsf{c} \right)$ is also always positive. 
We recall that we are considering the regime $r_2 \mathsf{c} + 2  r_1 r_2 - \mathsf{c}^2 < 0$ since otherwise, as we already proved, $\bar{\mathsf{C}}^{{\rm{HD}}}_{12} \geq \frac{1}{2} \mathsf{c}$. 
We have
\begin{align*}
&r_2 \mathsf{c} + 2  r_1 r_2 - \mathsf{c}^2 < 0 \stackrel{\text{from \ \eqref{eq:CutConstr1}}}{\Longrightarrow } 3 r_2 \mathsf{c} -2 r_2 \ell_2 -\mathsf{c}^2 <0 
\\& \Longrightarrow \ell_2 > \frac{3 r_2 \mathsf{c}-\mathsf{c}^2}{2 r_2}.
\end{align*}
Thus, we obtain
\begin{align*}
&\ell_2+\ell_3-\mathsf{c}
> 
\frac{3 r_2 \mathsf{c}-\mathsf{c}^2}{2 r_2} + \ell_3 - \mathsf{c}
\\&\stackrel{\text{from \ \eqref{eq:CutConstr2}}}{\geq} \frac{3 \mathsf{c} \left( \mathsf{c}-\ell_3 \right)- \mathsf{c}^2}{2 \left( \mathsf{c}-\ell_3 \right)} + \ell_3 - \mathsf{c} \!=\! \frac{\ell_3 \left( \mathsf{c}-2 \ell_3 \right)}{2 \left( \mathsf{c}-\ell_3\right)} \stackrel{\mathsf{c}\geq\ell_3 > \frac{\mathsf{c}}{2}}{>} 0.
\end{align*}
Thus, there are no channel conditions for which $\bar{\mathsf{C}}^{{\rm{HD}}}_{23}  <\frac{1}{2} \mathsf{c}$, i.e., by letting relay 2 and relay 3 operate we always achieve (to within a constant gap) at least half of the capacity of the corresponding FD network.
\\
{\bf{Case (iii-b):}}
$\ell_2 \ell_3 \geq r_2 r_3$; 
with this we obtain that the capacity $\bar{\mathsf{C}}^{{\rm{HD}}}_{23}$ of the pair relay 2 and relay 3 is given by \cite{BagheriIT2014}, i.e., 
\begin{align}
\bar{\mathsf{C}}^{{\rm{HD}}}_{23}  = \frac{\ell_2 \ell_3 r_3 - \ell_2 r_2^2 + \ell_2 r_2 r_3 + \ell_3 r_2 r_3}
{\left( \ell_2 + r_2\right) \left(\ell_3+r_3-r_2 \right)},
\end{align}
which implies 
$\frac{\bar{\mathsf{C}}^{{\rm{HD}}}_{23}}{{\mathsf{C}}^{{\rm{FD}}}_{\mathcal{N}_{\rm{F}}}} = \frac{\ell_2 \ell_3 r_3 - \ell_2 r_2^2 + \ell_2 r_2 r_3 + \ell_3 r_2 r_3}
{\mathsf{c}\left( \ell_2 + r_2\right) \left(\ell_3+r_3-r_2 \right)}$.
We now want to show that, in order to meet the conditions in \eqref{eq:CutConstr}, there are no networks for which
\begin{align*}
&\frac{\bar{\mathsf{C}}^{{\rm{HD}}}_{23}}{{\mathsf{C}}^{{\rm{FD}}}_{\mathcal{N}_{\rm{F}}}} = \frac{\ell_2 \ell_3 r_3 - \ell_2 r_2^2 + \ell_2 r_2 r_3 + \ell_3 r_2 r_3}
{\mathsf{c}\left( \ell_2 + r_2\right) \left(\ell_3+r_3-r_2 \right)} < \frac{1}{2},
\end{align*}
which is equivalent to show that there are no networks for which
\begin{align}
\label{eq:finalmente2}
&2 \ell_2 \ell_3 r_3 - 2 \ell_2 r_2^2 + 2 \ell_2 r_2 r_3 + 2 \ell_3 r_2 r_3 - \mathsf{c} \ell_2 \ell_3 - \mathsf{c} \ell_2 r_3 \nonumber
\\&+ \mathsf{c} \ell_2 r_2 -\mathsf{c} r_2 \ell_3 - \mathsf{c} r_2 r_3 + \mathsf{c} r_2^2 < 0.
\end{align}
It is not difficult to see that the LHS of \eqref{eq:finalmente2} is always increasing in $r_3$; the derivative of the LHS of \eqref{eq:finalmente2} with respect to $r_3$ is, in fact, given by
\begin{align*}
&2 \ell_2 \ell_3 + 2 \ell_2 r_2 + 2 \ell_3 r_2 - \mathsf{c} \ell_2 - \mathsf{c} r_2
\\& = \ell_2 \underbrace{\left( \ell_3 + r_2 -\mathsf{c} \right)}_{\geq 0 \ \text{from} \ \eqref{eq:CutConstr2}}
+ \underbrace{\ell_2 \ell_3 - \mathsf{c} r_2}_{\geq 0 \ \text{since} \ \ell_2 \ell_3 \geq r_2 r_3 \ \text{and} \ r_3 \geq \mathsf{c}}
\\&+\ell_2 r_2 +2\ell_3 r_2   \geq 0.
\end{align*}
Hence, from \eqref{eq:finalmente2} we obtain
\begin{align*}
&2 \ell_2 \ell_3 r_3 - 2 \ell_2 r_2^2 + 2 \ell_2 r_2 r_3 + 2 \ell_3 r_2 r_3 - \mathsf{c} \ell_2 \ell_3 - \mathsf{c} \ell_2 r_3 
\\& \quad + \mathsf{c} \ell_2 r_2 -\mathsf{c} r_2 \ell_3 - \mathsf{c} r_2 r_3 + \mathsf{c} r_2^2
\\ & \stackrel{r_3 \geq \mathsf{c}}{\geq}
\ell_2 \ell_3 \mathsf{c} - 2 \ell_2 r_2^2 + 3 \ell_2 r_2 \mathsf{c} + \ell_3 r_2 \mathsf{c} -\mathsf{c}^2 \ell_2 -\mathsf{c}^2 r_2 + \mathsf{c} r_2^2
\\ & = \mathsf{c} \left( \ell_2 +r_2 \right) \underbrace{\left( \ell_3 + r_2 - \mathsf{c}\right)}_{\geq 0 \ \text{from} \ \eqref{eq:CutConstr2}}
+ 2 \ell_2 r_2 \underbrace{\left( \mathsf{c} -r_2 \right)}_{\geq 0 \ \text{since} \ r_2 \leq \mathsf{c}} \geq 0.
\end{align*}
Thus, there are no channel conditions for which $\bar{\mathsf{C}}^{{\rm{HD}}}_{23} <\frac{1}{2} \mathsf{c}$, i.e., by letting relay 2 and relay 3 operate we always achieve (to within a constant gap) at least half of the capacity of the corresponding FD network. This concludes the proof of the lower bound in~\eqref{eq:PerfGarFD}.
We next provide a network example for which $\mathsf{C}^{{\rm{HD}}}_{2,3} \!=\! \frac{1}{2} {\mathsf{C}}^{{\rm{FD}}}_{\mathcal{N}_{\rm{F}}}$.
\\
{\bf{Example.}}
Consider
$\ell_1 \!=\! r_1 \!=\! 1, \ell_{[2:3]} \!=\! r_{[2:3]} \!=\! 0$;
clearly for this network we have ${\mathsf{C}}^{{\rm{FD}}}_{\mathcal{N}_{\rm{F}}} \!=\!1$ and $\mathsf{C}^{{\rm{HD}}}_{2,3} \!=\! \frac{\ell_1 r_1}{\ell_1+r_1}\!=\!\frac{1}{2}$; hence $\mathsf{C}^{{\rm{HD}}}_{2,3} \!=\!\frac{1}{2}{\mathsf{C}}^{{\rm{FD}}}_{\mathcal{N}_{\rm{F}}}$.

\bibliographystyle{IEEEtran}
\bibliography{isit2016}

\end{document}